\documentclass[11pt]{JHEP3}

\usepackage{amsmath,amsthm,amsfonts,amssymb,citesort,amscd,mathrsfs,graphicx,fontenc,Glebdefs,
}

\usepackage[normalem]{ulem}
\usepackage{color}
\let\normalcolor\relax
\usepackage{colordefs}

\newcommand{\mathcallig}[1]{ \mathbf{#1} }

\author{Gleb Arutyunov\footnote{E-mail: G.E.Arutyunov@uu.nl, M.deLeeuw@uu.nl, S.J.vanTongeren@uu.nl}
{}\footnote{Correspondent fellow at Steklov Mathematical
Institute, Moscow.}\,, \  Marius de Leeuw \, and\, Stijn J. van
Tongeren
 \\  {\it Institute for Theoretical Physics
and Spinoza Institute,\\ Utrecht University, 3508 TD Utrecht, The
Netherlands}}

\abstract{We study finite-size corrections to the magnon
dispersion relation in three models which differ from string
theory on $\AdS$ in their boundary conditions. Asymptotically,
this is accomplished by twisting the transfer matrix in a way
which manifestly preserves integrability. In model I all
world-sheet fields are periodic, whereas model II represents a particular orbifold
of $\AdS$ and model III is a $\beta$-deformed theory. For models I
and II we construct the one-particle TBA equations and use them to
determine the leading finite-size correction to the asymptotic
Bethe equation. We also make some interesting observations
concerning the quantization conditions for the momentum. For the
same models we compute the leading and for model II the
next-to-leading order finite-size corrections to the asymptotic
magnon dispersion relation. Furthermore, we apply L\"uscher's
formulae to compute the leading finite-size corrections in
$\beta$-deformed theory. In addition to reproducing known results,
we provide new predictions for two-particle states from the
$\sl(2)$ sector, to be confronted with explicit field-theoretic
calculations in the dual gauge theory. Finally, we prove that the
leading finite-size correction to the energy of an $\sl(2)$ magnon
in orbifold theory is the same as the one for an $\su(2)$ magnon
in $\beta$-deformed theory for special values of $\beta$. We also
speculate that for these values of $\beta$ our result for the
next-to-leading order correction in the orbifold model might
coincide with the corresponding correction to the energy of
$\su(2)$ magnon in $\beta$-deformed theory.}

\title{Twisting the Mirror TBA}

\preprint{
          \tiny{ITP-UU-10-33}\\[-.5ex]
          \tiny{SPIN-10-28}\\[-.5ex]
          }

\begin{document}
\renewcommand{\thefootnote}{\arabic{footnote}}
\setcounter{footnote}{0}

\section{Introduction}

In this paper we employ the mirror TBA in combination with
adaptations of L\"uscher's perturbative approach to study single
magnon configurations for various interesting models, which we discuss
in detail below. This has been made possible by  various important
recent developments in the problem of finding the exact scaling
dimensions of ${\cal N}=4$ SYM gauge-invariant composite operators
with finite quantum numbers through the gauge-string
correspondence \cite{Maldacena}.

\smallskip

First of all,  L\"uscher's perturbative
approach \cite{Luscher85} has been adapted to the case of the non-Lorentz invariant
\mbox{$\AdS$} string sigma-model \cite{JL07,BJ08}, which led to a
determination of the four-  and five-loop anomalous dimensions of
the Konishi operator \cite{BJ08,BJ09} and, subsequently, of all
twist two operators
\cite{Kotikov:2007cy,Bajnok:2008qj,Lukowski:2009ce}, see also
\cite{Beccaria:2009eq}; the corresponding results agree with
first-hand field-theoretic computations
\cite{Sieg,Vel,Fiamberti:2009jw}.

\smallskip

Secondly, the Thermodynamic Bethe Ansatz (TBA) equations \cite{AF09b,AF09d,Bombardelli:2009ns,GKKV09,Arutyunov:2009ax} for
the $\AdS$ mirror model\footnote{The TBA approach in the AdS/CFT
spectral problem was advocated  in \cite{AJK} where it was used to
explain wrapping effects in gauge theory.} \cite{AF07} have been constructed, based on
the corresponding string hypothesis \cite{AF09a}. These equations
should provide the tools needed to compute the energy of string
states and, therefore, the scaling dimensions of the corresponding operators in the dual gauge theory, as has been nicely demonstrated on the example of the
Konishi operator both at weak \cite{Arutyunov:2010gb,Balog:2010xa}
and intermediate \cite{GKV09b,Frolov:2010wt} values of the
coupling constant. The computation of the five-loop anomalous
dimension through the mirror TBA approach has been further
extended \cite{Balog:2010vf} to all twist two operators and the
results obtained agree with the ones based on the generalized
L\"uscher's formulae \cite{Lukowski:2009ce}. Finally, we note that
in general the TBA equations imply a set of functional equations
known as the Y-system \cite{ZamY}. The Y-system for the case at
hand  was conjectured in \cite{GKV09}, and its general solution
was obtained in \cite{Heg}. This Y-system has rather unusual
analytic properties which are under investigation
\cite{AF09b,FS,Arutyunov:2009ax,Cavaglia:2010nm}.

\smallskip

In the present paper we continue to explore the mirror TBA
approach. We concentrate on the TBA  describing a
one-particle excited state; a single magnon in the $\sl(2)$
sector with non-vanishing momentum $p$. This excitation does not
exist in the spectrum of closed strings on $\AdS$. The embedding
fields describing closed strings must satisfy periodic boundary
conditions which results in the level-matching condition; the
total momentum of a state must be equal to zero. Of course, the
concept of magnons remains of primary importance as it is used to
define their scattering matrix; a fundamental building block of
the whole integrability approach \cite{Zamolodchikov:1978xm}.

\smallskip

The spectral problem for a single magnon can be set up in
different ways. One way is to start from the string sigma-model
and {\it require} the embedding fields to be periodic. In this
model, which we call model I, we can study more excitations upon
relaxing the level-matching condition. Another possibility is to
consider strings in other integrable backgrounds, where the single
magnon is a physical excitation. Examples of this type are
provided by strings on orbifolds of $\AdS$ and $\beta$-deformed
theories.

\smallskip

In general our interest in the single magnon spectral problem is
three-fold. Firstly, we would like to understand the general
relationship between energy and momentum (dispersion) for
integrable models in a finite volume and its dependence on the
boundary conditions. Secondly, for theories where a single magnon
is a physical excitation, it is interesting to compute finite-size
corrections to the magnon energy, providing new data to be
compared to the corresponding dual gauge theories. Thirdly, it is
important to understand how to extend the TBA approach to a larger
class of physical theories.

\smallskip

In this paper we will analyze a one-particle excitation in {\it
three} different models. The first is the already mentioned model
I. Model II is defined by twisting the boundary conditions for
sphere bosons by the total momentum of a state. As we will show,
this model can be thought of as a particular orbifold of $\AdS$.
Finally, model III corresponds to $\beta$-deformed theory. All
models share the same world-sheet scattering matrix, which is also
the one for strings on $\AdS$
\cite{AFS,St04,Beisert:2005tm,Hernandez:2006tk,BHL06,BES,Bykov:2008bj}.
As a consequence, the modified boundary conditions for models II
and III in the large $J$ limit\footnote{Throughout the paper $J$
stands for the angular momentum of string rotating around the
equator of the five-sphere.} or at weak coupling  can be
implemented by means of the general theory of integrable twists
\cite{Sklyanin}. Below we will discuss the modified boundary
conditions starting from the underlying world-sheet theory.

\smallskip

For models I and II we construct the corresponding TBA equations.
Following the same strategy as \cite{Arutyunov:2009ax}, we assume
that the TBA equations for excited states only differ from  the
ground state ones by the integration contours in the convolution
terms. By requiring these equations to be compatible with the
large $J$ asymptotic solution given in terms of generalized
L\"uscher formulae we then fix the shape of the integration
contours. Simultaneously, we also determine how the TBA length
parameter $L_{\scriptsize TBA}$ is related to $J$. Interestingly
enough, for model I there is a noticeable difference between the
level-matched and unmatched cases; $L_{\scriptsize TBA}=J+2$ in
the first case and $L_{\scriptsize TBA}=J$ in the second. For
model II we find an intermediate result for the length;
$L_{\scriptsize TBA}=J+1$.

\smallskip

Furthermore, by analytically continuing the TBA equation for the
$Y_1$-function to the kinematical region of string theory along
the lines of \cite{DT96}, we derive an explicit form of the exact
Bethe equation $Y_1(u_*)+1=0$ for the rapidity variable $u_*$.
This equation should be viewed as a {\it quantization condition}
for the particle momentum (rapidity) which replaces the asymptotic
Bethe equation for the case of finite $J$.

\smallskip

The large $J$ asymptotic solution is constructed through L\"uscher
formulae in terms of the transfer matrices in which the boundary
conditions of interest are implemented through the corresponding
twists. The L\"uscher formulae allow us to compute the leading
order finite-size corrections to the energy $E_{\rm LO}$ and
momentum $P_{\rm LO}$. As was found for relativistic integrable
models starting from an underlying lattice formulation, see
example \cite{Balog:2003xd,Teschner:2007ng}, the energy $E$ and
momentum $P$ are given by equations (\ref{FSE}) and
(\ref{standardmomentum}), respectively. In the case of non periodic
boundary conditions, {\it i.e.} for models II and III, based on
experience with the dual gauge theories, it is plausible to assume
that the expression for the asymptotic energy is not modified. The
situation regarding $P$ is less clear however, and we just adopt
formula (\ref{standardmomentum}) as a working assumption. We then
show that for model I the leading finite-size correction to the
asymptotic Bethe equation coincides with $P_{\rm LO}$, while for
model II this is not the case. A possible reconciliation of the
two is that the introduction of a twist modifies the definition of
the momentum in such a way that it becomes compatible with the
exact Bethe equation.

\smallskip

For model II, by employing some of the techniques of \cite{BJ09},
we have computed the next-to-leading order correction to the
energy $E_{\rm NLO}$ of an $\sl(2)$ magnon for $J=2,3,4$ and $6$\footnote{Here the results take a nicely presentable form.}.
The results are given in terms of $\zeta$-functions and products
thereof; all polygamma-functions arising in various steps of the
computation are canceled in the final expressions. In these
energy corrections, the term of maximal transcendentality has
degree $2J+3$. We also present results for the energy correction
to some two-particle states from the $\sl(2)$ sector for a generic
${\mathbb Z}_S$-orbifold. As opposed to the single magnon, these
states satisfy the level-matching condition, {\it i.e.} they are physical states of the orbifold model.

\smallskip

Concerning $\beta$-deformed theory, the construction of the TBA
equations remains an open problem. Moreover, there is no consensus
on the question of integrability of the finite-size
model\footnote{For instance, the equations which encode a class of finite-gap solutions of the $\gamma$-deformed string sigma model have been discussed in \cite{Frolov:2005ty}. However, these equations have been derived under an assumption that certain non-linear constraints are satisfied (in the limit $\gamma \rightarrow 0$ these constraints turn into the global conserved charges that do not belong to the Cartan subalgebra of $\rm{SO}(6)$). It is unclear for the moment how to extend the corresponding approach to encompass all finite-gap solutions. We thank Sergey Frolov for the discussion of this point.}. Here we restrict ourselves to the study of the leading
L\"uscher corrections to the energies of states in the $\su(2)$
and $\sl(2)$ sectors. The L\"uscher formulae assume factorization
of the full world-sheet S-matrix into a product of two
$\su(2|2)$-invariant S-matrices, and in this aspect they still
reflect the properties of the large $J$ solution. Understanding
double wrapping should provide crucial insight on the issue of
integrability of the $\beta$-deformed theory in finite volume.

\smallskip

Recently, wrapping energy corrections were found in $\beta$-deformed SYM \cite{Fiamberti:2008sm,Fiamberti:2008sn}, which sparked attempts to reproduce these from finite size effects in the dual string theory. Close to what we consider here, an approach of obtaining the leading finite-size
corrections to the energies of $\su(2)$ states in $\beta$-deformed
theories via twisting has been implemented in \cite{Gromov:2010dy}. However, the twist we are using is slightly different from the one in \cite{Gromov:2010dy} and we justify it from the
world-sheet point of view. At the same time, our transfer matrix
is also different from the one in \cite{Gromov:2010dy}, because it
contains extra momentum-dependent prefactors needed to satisfy the
generalized unitarity condition \cite{AF07}. As a net result, the
$Y_Q$-functions in the $\su(2)$ sector agree with the ones found in \cite{Gromov:2010dy} for any number of excitations, and therefore in particular the L\"uscher corrections obtained there for a single $\su(2)$ magnon coincide with ours. Additionally, in \cite{Ahn:2010yv} the $\beta$-deformed
analog of the Konishi state was studied by means of twisting
certain elements of the S-matrix. Our twist incorporates this
result, but derives it from a different starting point.

\smallskip

We then turn our attention to the $\sl(2)$ sector in
$\beta$-deformed theory with $\beta\neq \frac{n}{J}$, where $n$ is
an integer. We determine the leading order corrections to the
energy of two-particle states with $J=2,3,5$ and give explicit
formulae that can be readily used to compute energy corrections
for any $J$ desired, providing new predictions to be confronted
with future calculations in the dual gauge theory. We point out
that for rational values $\beta=\frac{n}{J}$ the correction starts
at an order of  $g^2$ higher than for generic $\beta$.

\smallskip
We also make a detailed comparison of orbifold and
$\beta$-deformed theories and show that for $\beta=\frac{n}{J}$
the leading order finite-size correction to the energy of an $\sl(2)$
magnon in the orbifold model is the same as the corresponding
quantity for an $\su(2)$ magnon in $\beta$-deformed theory, in full
agreement with the conjecture by \cite{Gunnesson:2009nn}. We point
out that if this interesting relation between the two theories
continues to hold beyond the leading order, our result for
$E_{\rm NLO}$ found for the orbifold model will simultaneously give the NLO
correction to the energy of an $\su(2)$ magnon in $\beta$-deformed
theory for these special values of $\beta$.

\smallskip

The paper is organized as follows. In the next section we
introduce the three models and discuss implementation of the
corresponding twists at the level of transfer matrices. Section 3
contains some generalities on energy and momentum in the TBA
approach. Section 4 is devoted to the construction of the TBA
equations for models I and II. In sections 5 and 6 we compute the
leading finite-size corrections to the energy and momentum in
models I and II, respectively. Section 6 also contains some
results on the next-to-leading order correction to the energy of
the $\sl(2)$ magnon in model II. Section 7 contains our results
for $\beta$-deformed theories. Some definitions and technical
details are relegated to appendices A, C and D. In appendix B we
briefly discuss the issue of critical points for the asymptotic
solution of the TBA equations.

\smallskip

We note here that sections 2 and 3 are of general interest, which
combined with section 6 contain the background and explicit
results on the leading order energy corrections in the orbifold
theory, while combined with section 7 they contain the background
and explicit results for $\beta$-deformed theory and two particle $\alg{sl}(2)$ orbifold states. Sections 4, 5
and 6 contain the results which are important for next-to-leading
order (and higher) corrections and our observations regarding
momentum quantization.

\section{Models: periodic, orbifold and $\beta$-deformed}  An important peculiarity of the
string sigma-model on $\AdS$ is that the fermions exhibit unusual
periodicity conditions related to the total momentum $p$
\cite{Alday:2005jm,Arutyunov:2009ga}. Indeed, in the light-cone
gauge one of the fields, an angle $\phi$ which parametrizes the
five-sphere, is unphysical and must be solved in terms of physical
(transversal) fields. The equation of motion for $\phi$ then
implies \bea \phi(2\pi)-\phi(0)=p\, . \label{eomphi}\eea
 On the other hand, in order to make the
world-sheet fermions neutral under the isometry $\phi\to\phi+{\rm
constant}$, and simultaneously bring the Wess-Zumino term of the
string action to a local form, we need to perform a  field
redefinition of the fermions $\psi\to e^{\frac{i}{2}\phi}\psi$. As a
result, the new fermions satisfy the {\it twisted} boundary
conditions \bea \label{tf} \psi(2\pi)=e^{\frac{i}{2}p}\psi(0)\, .
\eea Since we aim to describe closed strings, all the sphere
bosons must be periodic which leads to the level-matching
condition $p=2\pi m$, where $m$ is the winding number. As a
result, the theory splits into sectors with periodic and
anti-periodic boundary conditions for fermions, depending on whether
$m$ is even or odd.

\smallskip

Moving on, we recall that in the large $J$ limit the symmetry algebra of the string
sigma-model coincides with two copies of
the centrally extended $\su(2|2)$ superalgebra with the same
central charge depending on $p$
\cite{Beisert:2005tm,Arutyunov:2006ak}. The fact that the 16 physical
fields (8 bosons and 8 fermions) transform under two copies of
$\su(2|2)$ suggests to treat this representation as a tensor
product $4\times 4$, where $4$ is a fundamental four-dimensional
representation of $\su(2|2)$ with two bosons and two fermions.

\smallskip

In order to treat a single magnon, we have to relax the
level-matching condition. It is then easy to see that it is
impossible to twist the boundary conditions for factorized
world-sheet fields in such a way as to have all sphere and AdS
bosons strictly periodic, while fermions satisfying the twisted
boundary conditions (\ref{tf}). In the following we therefore
introduce three models which have the same field content as the
string sigma-model on $\AdS$ but differ in their boundary
conditions.  Within these models we will be able to study the
spectral problem for a single magnon.

\subsubsection*{Model I -- Periodic fields} In this model one assumes that all
fields -- bosons and fermions -- are periodic. The level-matching
condition is not imposed allowing to consider a single magnon with
non-vanishing momentum. Clearly, the periodic boundary conditions
preserve all the global symmetries of the original light-cone
string sigma-model.

\vskip 0.3cm

\subsubsection*{Model II -- Orbifold} In this model one twists the boundary
conditions for bosonic fields $y_{a}$ and $y_{\dot{a}}$, where
$a=1,2$ and $\dot{a}=\dot{1},\dot{2}$: \bea\nonumber
\left(\begin{array}{c} y_{1}(2\pi) \\ y_{2}(2\pi)
\end{array} \right)=\left(\begin{array}{cc}
e^{i\alpha_{\ell}} & 0 \\ 0 & e^{-i\alpha_{\ell}}
\end{array} \right)\left(\begin{array}{c}
y_{1}(0) \\ y_{2}(0)
\end{array} \right)\,
, ~~~~~~ \left(\begin{array}{c} y_{\dot{1}}(2\pi) \\
y_{\dot{2}}(2\pi)
\end{array} \right)=\left(\begin{array}{cc}
e^{i\alpha_r} & 0 \\ 0 & e^{-i\alpha_r}
\end{array} \right)\left(\begin{array}{c}
y_{\dot{1}}(0) \\ y_{\dot{2}}(0)
\end{array} \right),\eea while fermions $\theta_{\a}$ and $\theta_{\dot{\alpha}}$
kept untwisted and so the AdS fields as well. Here we should
distinguish two possibilities for choosing the left and right
twists: \bea\label{2twists} \alpha_{\ell}=\alpha_r~~~~~{\rm
or}~~~~~\alpha_{\ell}=-\alpha_r\, . \eea
The twists above will lead to a modification of the boundary conditions for two of the four fields $Y_{a\dot{a}}$. Together with the two fields contained in the complex $Z$ field which is necessarily unaffected by the twist, these fields parametrize the five-sphere. Furthermore the physical
fermions $\theta_{a\dot{\alpha}}=y_a\theta_{\dot{\alpha}}$ and
$\eta_{\dot{a}\alpha }=\theta_{\alpha}y_{\dot{a}}$ inherit
the twisted periodicity conditions \bea \label{tbcf}
\begin{aligned}
&\theta_{1\dot{\alpha}}(2\pi)=e^{i\alpha_{\ell}}\theta_{1\dot{\alpha}}(0)\,
, &~~~~~ &\eta_{ \dot{1}\alpha}(2\pi)=e^{i\alpha_{r}}\eta_{\dot{1}\alpha}(0)\, , \\
&\theta_{2\dot{\alpha}}(2\pi)=e^{-i\alpha_\ell}\theta_{2\dot{\alpha}}(0)\,
, &~~~~~ &\eta_{ \dot{2}\alpha}(2\pi)=e^{-i\alpha_r}\eta_{\dot{2}\alpha}(0)\, .
\end{aligned}\eea
The two twists in equation (\ref{2twists}) lead to essentially equivalent
theories, so in what follows we have chosen to explicitly treat
the case $\alpha_{\ell}=-\alpha_r$.

\smallskip

A particularly interesting situation arises when a twist is chosen
to be \bea\label{orbi}\alpha_{\ell}=\frac{2\pi n}{S}\, ,\eea where
$n$ and $S$ are integers. In this case the twisted boundary
conditions coincide with the ones for string theory on the
orbifold $\AdS/{\Gamma}$, where $\Gamma$ is a cyclic subgroup
$\mathbb{Z}_{S}$ in $\su(2)\subset\su(4)$, see
\cite{Ideguchi:2004wm,Beisert:2005he}. The canonical element
generating $\mathbb{Z}_{S}$ is $\omega=e^{\frac{2\pi i }{S}}$ and
various values of $n$, $n=0,\ldots , S-1$ describe the
corresponding twisted sectors. Since two of the three complex
scalars, $Z$ and $Y_{1\dot{1}}$, do not undergo twisting, the
orbifold model preserves an $\su(2)$ part of the original $\su(4)$
R-symmetry on the bosonic fields. However, since all fermions are twisted, it has no supersymmetry\footnote{We would like to thank Matteo Beccaria for correcting a statement about the presence of supersymmetry in this model in the previous version of this paper.}.

\smallskip

Yet another possibility, which we will be mostly interested in, is
to chose $\alpha_{\ell}=-\alpha_{r}=\frac{P}{2}$, where $P$ is the
total momentum of a state. Excitations in this model with these
boundary conditions are embedded in the orbifold model. Indeed, at
large $J$ the momentum of a single magnon is subjected to the
quantization condition $p=\frac{2\pi n}{J}$, $n\in\mathbb{Z}$,
which is nothing else but the Bethe equation, {\it cf.} our
discussion at the end of this section. Therefore, for fixed $J$
we deal with a magnon in the twisted sector corresponding to $\omega^n=e^{\frac{i\pi
n}{J}}$ of the orbifold theory. As we will see, the momentum $p$ receives finite-size corrections starting at
order $g^{2J+4}$; these effects are compatible with the quantization condition, and hence preserve the orbifold structure. For LO and NLO corrections to the energy of a single magnon the difference between $p$ and the
total momentum will not play any role.

\smallskip

In both model I and II we will study states from the
$\sl(2)$ sector. For states from this sector the level-matching
condition implies that the total momentum of a state must be equal
to zero. In particular therefore a single $\sl(2)$ magnon is not a physical
state of the orbifold theory. In what follows we will therefore
study one-magnon configurations by relaxing the level-matching
condition. In section 7 we also compute the leading finite-size
corrections for a few {\it physical} two-particle states of a
generic orbifold.

\smallskip

It is worth pointing out that the orbifold boundary conditions with the twist parameter chosen to
be proportional to the total momentum introduced here, are similar but not equivalent to those of
$\beta$-deformed theories. Nevertheless, as will show this model exhibits an interesting relation to  $\beta$-deformed theories.

\subsubsection*{Model III -- $\beta$-deformations}

This is a model which describes an exactly marginal deformation of
$\AdS$ superstring theory. It has been introduced in
\cite{Lunin:2005jy} and further studied in
\cite{Frolov:2005dj,Beisert:2005if,Frolov:2005iq,Alday:2005ww,Frolov:2005ty}. The model admits a natural generalization to a more general
non-supersymmetric background, which is obtained as a
three-parameter deformation of $\AdS$. The latter can be regarded
as the same $\AdS$ string sigma-model but with twisted boundary
conditions for three isometric angles $\phi_i$ of ${\rm S}^5$,
namely, \bea\hspace{1cm} \label{deform}
\phi_i(2\pi)-\phi_i(0)=-2\pi\epsilon_{ijk}\gamma_jJ_k \,
,~~~~~~i=1,2,3.\eea In these $\gamma$-deformed theories, the
$\gamma_i$ are three deformation parameters and the  $J_i$ are
conserved Noether charges corresponding to the shift isometries in the $\phi_i$ direction. Because of the presence of the $J_i$, the deformations depend on
the state of interest. For the symmetric case \bea
\gamma_1=\gamma_2=\gamma_3=\beta \label{betacase}\eea with $\beta$
real, the deformation preserves 8 supercharges and it is dual to
the ${\cal N}=1$ superconformal field theory which is an exactly
marginal deformation of ${\cal N}=4$ super Yang-Mills theory
\cite{Lunin:2005jy}.

\smallskip

To continue, let us recall that the transverse bosonic fields describing the
five-sphere can be combined in a matrix $Y$ \bea
Y=\left(\begin{array}{cc} Y_{1\dot{1}} ~&~ Y_{1\dot{2}} \\
Y_{2\dot{1}} ~&~ Y_{2\dot{2}}
\end{array}\right)= \left(\begin{array}{cc} ae^{-i\phi_2} ~&~ be^{-i\phi_3} \\
be^{i\phi_3} ~&~ -ae^{i\phi_2}
\end{array}\right)\, ,\eea
where we exhibit explicitly the dependence on two isometric angles
$\phi_2$ and $\phi_3$ \cite{Arutyunov:2009ga}. The other angle
$\phi_1$ parametrizes the complex field
$$
Z=\frac{1-a^2-b^2}{1+a^2+b^2}e^{i\phi_1}\, .
$$
Thus, the three $\phi$'s together with $a$ and $b$ provide a parametrization of
the five-sphere. We see that equation (\ref{deform}) implies the
following twisted boundary conditions for sphere bosons \bea
Y(2\pi) =\left(
\begin{array}{cc} e^{i\pi\beta(J_2-J_3)} & 0 \\
0 & e^{-i\pi\beta(J_2-J_3)}
\end{array}\right)Y(0)\left(\begin{array}{cc} e^{i\pi\beta(2J_1-J_2-J_3)} & 0 \\
0 & e^{-i\pi\beta(2J_1-J_2-J_3)}
\end{array}\right) \, .\eea
  Obviously,
these conditions are inherited from the following periodicity
conditions for $y_{a}$ (left bosons) and $y_{\dot{a}}$ (right
bosons) \bea \label{LR}
 \left(\begin{array}{c} y_1(2\pi) \\
y_2(2\pi)
\end{array}\right)&=&\left(\begin{array}{cc} e^{i\pi\beta(J_2-J_3)} & 0 \\
0 & e^{-i\pi\beta(J_2-J_3)}
\end{array}\right)\left(\begin{array}{c} y_1(0) \\
y_2(0)
\end{array}\right)\, ,
\\
 \left(\begin{array}{c} y_{\dot{1}}(2\pi) \\
y_{\dot{2}}(2\pi)
\end{array}\right)&=&\left(\begin{array}{cc} e^{i\pi\beta(2J_1-J_2-J_3)} & 0 \\
0 & e^{-i\pi\beta(2J_1-J_2-J_3)}
\end{array}\right)\left(\begin{array}{c} y_{\dot{1}}(0) \\
y_{\dot{2}}(0)
\end{array}\right)\, .
\eea Also, we have \bea Z(2\pi)=Z(0)e^{2\pi \beta(J_2-J_3)}\, .
\eea As was mentioned above, in the light-cone gauge the field
$\phi\equiv \phi_1$ is non-dynamical and satisfies
equation (\ref{eomphi}). Thus, $p=2\pi \beta(J_2-J_3)$ is the
level-matching condition in $\beta$-deformed theory. Finally, we
mention that the fermions $\theta_{\alpha}$ and
$\theta_{\dot{\alpha}}$ have standard periodic boundary
conditions, while the physical fermions
$\theta_{a\dot{\alpha}}=y_a\theta_{\dot{\alpha}}$ and
$\eta_{\alpha \dot{a}}=\theta_{\alpha}y_{\dot{a}}$ will inherit
the twisted boundary conditions from the bosons.

\smallskip

In $\beta$-deformed theory we will be interested in two kind of
states which we call states from the $\su(2)$ and $\sl(2)$
sectors respectively. The states from $\su(2)$ have $J_3=0$ and
are characterized by two charges $J\equiv J_1$ and $M\equiv J_2$;
the level-matching condition is $p=2\pi\beta M$. For $J$ large
such a state can be represented as
$$
\underbrace{A_{1\dot{1}}^+(p_1)\ldots A_{1\dot{1}}^+(p_M)}_{M} |0 \rangle \, ,
$$
where $A_{1\dot{1}}^+$ is a (bosonic) creation operator from the
corresponding Faddeev-Zamolodchikov algebra
\cite{Arutyunov:2006yd}. The gauge theory operators dual to the
states from $\su(2)$ have schematically the form ${\rm tr}(\Phi^M
Z^J)$, where a complex scalar $\Phi$ is dual to $Y_{1\dot{1}}$. A
single $\su(2)$ magnon with $p=2\pi \beta$ is a {\it physical}
excitation of $\beta$-deformed theory.

\smallskip

The states from $\sl(2)$ have a single charge $J$ and a spin $S$,
and in the large $J$ limit can be represented as
$$
\underbrace{A_{3\dot{3}}^+(p_1)\ldots A_{3\dot{3}}^+(p_S)}_{S}|0 \rangle \, ,
$$
where $3$ and $\dot{3}$ are fermionic indices. Since $J_2=M=0$,
the level matching condition\footnote{For simplicity we restrict
ourselves to the case of no winding.} is $p=0$. A single
$\sl(2)$ magnon can be considered off-shell, but in this case it
is not a {\it physical} excitation of $\beta$-deformed theory. The
dual gauge theory operators are schematically ${\rm tr}(D^S Z^J)$.

\subsubsection*{Twisted transfer matrix}
In the large $J$ limit the spectrum of the models discussed above is
captured by the asymptotic Bethe Ansatz equations (the Bethe-Yang
equations) which are constructed taking into account the corresponding
boundary conditions. In the framework of the Algebraic Bethe
Ansatz, (modified) boundary conditions are accounted for by means of twisting
the transfer matrix. Now consider $M$ string theory particles
characterized by the $u_*$-plane rapidities $u_1,\ldots, u_M$ or,
equivalently, with momenta $p_1,\ldots,p_M$. Consider also an
auxiliary particle with rapidity $v$ corresponding to a bound
state representation $\pi_Q$ of $\su(2|2)$ with the bound state
number $Q$. Scattering this auxiliary particle through $N$
particles gives rise to a monodromy matrix
$$
\mathbb{T}(v|\vec{u})=\prod_{i=1}^M \mathbb{S}_{ai}(v,u_i)\, .
$$
Here $\mathbb{S}_{ai}(v,u_i)$ is the S-matrix which describes
scattering of the auxiliary particle with another one with
rapidity $u_i$. As a matrix acting on the auxiliary space,
$\mathbb{T}(v|\vec{u})$ satisfies the fundamental commutation
relations
$$
\mathbb{S}_{12}(v_1,v_2)\mathbb{T}_1(v_1|\vec{u})\mathbb{T}_2(v_2|\vec{u})=\mathbb{T}_2(v_2|\vec{u})\mathbb{T}_1(v_1|\vec{u})
\mathbb{S}_{12}(v_1,v_2)\, .
$$
We can also introduce a {\it twisted transfer matrix} \bea T(v|\vec{u})={\rm
Tr}\Big[\pi_Q(g)\, \mathbb{T}(v|\vec{u})\Big] \, ,\eea where an
element  $g$ is a twist and trace is taken over the auxiliary
space. If $g$ is such that $[\mathbb{S}_{12},g\otimes g]=0$, then
the fundamental commutation relations imply that $T(v|\vec{u})$
commute for different values of $v$ and therefore define a set
of commuting charges. For the string S-matrix we are interested
in  $g\in {\rm SU(2)}\times {\rm SU(2)}$ and as such the most
general twist will involve six arbitrary parameters. Here the
first ${\rm SU(2)}$ corresponds to the AdS space, while the second
one corresponds to the five-sphere.

\smallskip

Postponing development of the general theory to the future\footnote{ A
generic twist corresponds to describing the most general orbifold
and $\gamma$-deformed theories; it breaks global symmetries of the
model and brings the Bethe roots at infinity to finite positions
in the complex plane \cite{Frolov:2005iq,Bazhanov:2010ts}. }, we
note that for all cases of present interest the twist is of
the form $g=1\otimes K$, where \bea K=\left(\begin{array}{cc} e^{i\alpha} & 0 \\
0 & e^{-i\alpha}
\end{array}\right)\, .\eea

\begin{itemize}
\item For model I $\alpha=0$, {\it i.e.} twist is absent for both
left and right copies of $\su(2|2)$.

\item For model II we have two options  \bea
\alpha_{\ell}=\alpha_r=\frac{p}{2}~~~~~{\rm
or}~~~~~\alpha_{\ell}=-\alpha_r=\frac{p}{2}\, , \eea for the left
and $\alpha_{\ell}$ and $\alpha_r$ refer to the left and the right
transfer matrices, respectively.

\item For $\beta$-deformed theory the left and right
transfer matrices corresponding to left and right copies of
$\su(2|2)$ are twisted with \bea \alpha_{\ell}=\pi\beta(J_2-J_3)\,
, ~~~~\alpha_{r}=\pi\beta(2J_1-J_2-J_3)\,  \eea for the left and
the right transfer matrices, respectively.
\end{itemize}

\begin{table}
\begin{center}
\begin{tabular}{|p{2.8cm} | p{2cm} | p{2cm} | p{2cm} | p{2cm} |}
\hline
Sector& \multicolumn{2}{|c|}{$\alg{sl}(2)$} & \multicolumn{2}{|c|}{$\alg{su}(2)$}\\
\hline
Factor & L & R & L & R \\
\hline \hline
Periodic (I) & 0 & 0 & 0 & 0 \\
\hline
Orbifold (II) & $p/2$  &  $\pm p/2$ & $p/2$ & $\pm p/2$ \\
\hline
$\beta$-deformed (III) & 0 &  $2\pi J\beta$ &  $\pi M\beta$  & $(2J-M) \pi \beta$ \\
\hline
\end{tabular}
\caption{The twists for the different models in the
$\alg{sl}(2)$ and $\alg{su}(2)$  sectors.  The off-shell
$\alg{su}(2)$ magnon with zero twist has not been investigated.}
\label{tab:twists}
\end{center}
\end{table}

\noindent For the reader's convenience we have summarized the twists for the different models in the $\sl(2)$ and $\su(2)$ sectors in Table \ref{tab:twists}.

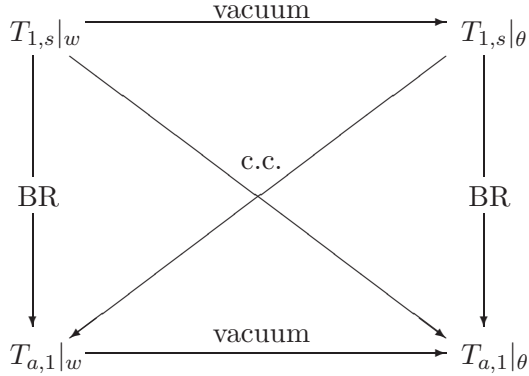
\begin{figure}
\begin{center}
{ \setlength{\unitlength}{1mm}
\begin{picture}(80, 50)
\put(0,43){$T_{1,s}|_w$} \put(3, 41){\line(0, -1){16}} \put(1,
21){\text{BR}} \put(3, 20){\vector(0, -1){15}} \put(10,
45.5){\vector(1, 0){48}} \put(0,0){$T_{a,1}|_w$} \put(10,
1.5){\vector(1, 0){48}} \put(60,0){$T_{a,1}|_\theta$} \put(63,
41){\line(0, -1){16}} \put(61, 21){\text{BR}} \put(63,
20){\vector(0, -1){15}} \put(60,43){$T_{1,s}|_\theta$} \put(8,
41){\vector(4,-3){50}} \put(58, 41){\vector(-4,-3){50}}
\put(30.7,26){\text{c.c.}} \put(27,3){\text{vacuum}}
\put(27,46){\text{vacuum}}
\end{picture}
}
\end{center}
\caption{ The relations between eigenvalues of the transfer matrix
calculated for (anti-)symmetric representations on a bosonic ($w$)
and fermionic ($\theta$) vacuum \cite{Arutyunov:2009pw}. BR stands
for the Bazhanov-Reshetikhin formula \cite{BR}, c.c. for complex
conjugation or equivalently for switching the scattering matrix
$S$ for $S^{-1}$.}

\label{fig:Trelations}
\end{figure}

\vskip 0.5cm The $Q$-particle bound state representations of
$\su(2|2)$ have dimension $4Q$ and are of two types;
symmetric and anti-symmetric. Hence we can define two
transfer matrices, $T_{1,s}$ and $T_{a,1}$ corresponding to
symmetric and anti-symmetric bound state representations
respectively. The relations between $T_{1,s}$ and $T_{a,1}$
are schematically depicted in Figure \ref{fig:Trelations}. These transfer matrices
are now used to construct the asymptotic Y-functions $Y_Q$ in the
$\sl(2)$ and $\su(2)$ sectors\footnote{The
reader should not confuse the asymptotic $Y_Q^{o}$- and exact
$Y_Q$-functions. Since we are mainly interested here in the
asymptotic solution, we will often omit the superscript $``o"$,
for instance, in section 4 where the TBA equations are
discussed.}. The corresponding expressions are given by the
generalized L\"uscher formula \cite{BJ08}; \bea\label{genericY}
Y^{o}_Q(v)=e^{-J\tilde{\cal
E}_Q(v)}T^{\ell}(v|\vec{u})T^{r}(v|\vec{u})\prod_{i=1}^M
S_{\sl(2)}^{Q1_*}(v, u_i) \, .\eea Here $\tilde{\cal E}_Q$ is the
energy of a mirror $Q$-particle, $S_{\sl(2)}^{Q1_*}(v,u)$ denotes
the S-matrix with the first and second argument in the mirror
and string regions, and $T^{\ell,r}$ is the left, respectively right
twisted $\su(2|2)$ transfer matrix.

\smallskip

Below we discuss the asymptotic Bethe equations which in the
present approach arise from the condition $Y_{1*}(u_j)=-1$, the
latter being the equations for the string rapidities $u_j$. Here
$Y_{1*}$ is the asymptotic $Y_1(v)$ function analytically
continued to the kinematic region of string theory.

\subsection*{Bethe equations in the $\sl(2)$ sector}
In appendix A we present the full eigenvalues of the twisted
transfer matrix $T_{Q,1}$ evaluated on the $\alg{sl}(2)$ vacuum, as well as the corresponding auxiliary
Bethe equations which can be easily obtained by generalizing the
result of \cite{Arutyunov:2009iq} found for the untwisted case.
For states from the $\sl(2)$ sector we have
$T^{\ell}=T^{r}=T_{Q,1}^{\alg{sl}(2)}$, where the eigenvalue $T_{Q,1}^{\alg{sl}(2)}$ does not
involve auxiliary Bethe roots and is given by
 \bea\label{TS}
T_{Q,1}^{\alg{sl}(2)}(v\,|\,\vec{u})&=&1+\prod_{i=1}^{M} \frac{(x^--x^-_i)(1-x^-
x^+_i)}{(x^+-x^-_i)(1-x^+
x^+_i)}\frac{x^+}{x^-}\\
&&\hspace{-1.5cm}-2\cos\alpha\sum_{k=0}^{Q-1}\prod_{i=1}^{M}
\frac{x^+-x^+_i}{x^+-x^-_i}\sqrt{\frac{x^-_i}{x^+_i}}
\left[1-\frac{\frac{2ik}{g}}{v-u_i+\frac{i}{g}(Q-1)}\right]+\sum_{m=\pm}
\sum_{k=1}^{Q-1}\prod_{i=1}^{M}\lambda_m(v,u_i,k)\, . \nonumber
\eea Definitions of various quantities entering the last formula
can be found in appendix A; $\cos{\alpha}$ is a twist of the bosonic
eigenvalues. The subscript of the transfer matrix indicates the
anti-symmetric bound state representation which an auxiliary
particle transforms in upon continuation of its spectral parameter
from the mirror to string theory kinematical region. The transfer
matrix is normalized as $T_{1*,1}^{\alg{sl}(2)}=1$, so that the asymptotic Bethe
equations in the $\sl(2)$ sector have the form \bea\label{BYsl2}
1=e^{ip_j J}\prod_{i\neq j}^M S_{\sl(2)}(u_j, u_i) \, .\eea Here
$S_{\sl(2)}^{1_*1_*}(u_j, u_i)\equiv S_{\sl(2)}(u_j, u_i)$ is the
usual $\sl(2)$ sector S-matrix. In particular, for a single
excitation this gives
\begin{equation*}
1=e^{i p J}.
\end{equation*}
From here, note that in none of the models we consider the twist enters
the $\sl(2)$ Bethe equations.  This of course should be the case
because in all our models, including the $\beta$-deformed one, the
$\sl(2)$ bosons have usual periodic boundary conditions.

\subsection*{Bethe equations in the $\su(2)$ sector}
In the $\su(2)$ sector the basic object needed to build up the $Y_Q$ functions through the
generalized L\"uscher formulae is the transfer matrix $T_{Q,1}^{\alg{su}(2)}$,
the left and right copies of which are generically twisted
differently. For the twisted $T_{Q,1}^{\alg{su}(2)}$ one has
 \bea
T_{Q,1}^{\alg{su}(2)}&=&(Q+1)\prod_{i=1}^{M}\frac{x^--x_i^-}{x^+-x_i^-}\sqrt{\frac{x^+}{x^-}}
-Q\, e^{-i \alpha}
\prod_{i=1}^{M}\frac{x^--x_i^+}{x^+-x_i^-}\sqrt{\frac{x^+x_i^-}{x^-x_i^+}}
-\\ \nonumber &&-Q\, e^{i\alpha}
\prod_{i=1}^{M}\frac{x^--x_i^-}{x^+-x_i^-}\frac{x_i^--\frac{1}{x^+}}{x_i^+-\frac{1}{x^+}}
\sqrt{\frac{x^+x_i^+}{x^-x_i^-}}+(Q-1)\prod_{i=1}^{M}\frac{x^--x_i^+}{x^+-x_i^-}\frac{x_i^--\frac{1}{x^+}}{x_i^+-\frac{1}{x^+}}
\sqrt{\frac{x^+}{x^-}}\, . \eea This time the subscript of the
transfer matrix indicates the symmetric bound state representation
which an auxiliary particle transforms in upon continuation of its
spectral parameter from the mirror to string theory kinematical
region. The asymptotic Bethe equations now read \bea\label{BYsu2}
-1=Y_{1_*}(u_k)= e^{ip_k
J}e^{-i(\alpha^{\ell}+\alpha^{r})}\prod_{i=1}^M S_{\sl(2)}(u_k,
u_i)\left(\frac{x_k^--x_i^+}{x_k^+-x_i^-}\sqrt{\frac{x_k^+x_i^-}{x_k^-x_i^+}}\right)^2\,
. \eea In particular, for a single excitation we
have\footnote{Recall that $S_{\sl(2)}(u, u)=-1$.} \bea 1=e^{i(p
J-\alpha_{\ell}-\alpha_r)}\, . \label{1magnonsu2}\eea

\section{Energy and Momentum}
In the TBA approach for the $\AdS$ mirror model, the energy of an
$M$-particle state from the $\sl(2)$-sector is given by
\begin{align}
\label{FSE} E =J+\sum_{i=1}^M{\cal E}(p_i)
-\frac{1}{2\pi}\sum_{Q=1}^{\infty}\int {\rm d}v
\frac{d\tilde{p}^Q}{dv}\log(1+Y_{Q}(v)).
\end{align}
Here the integration runs over a real rapidity line of the mirror
theory and $\tilde{p}^Q$ are momenta of the mirror $Q$-particles.
Moreover, ${\cal E}(p)$ is the asymptotic energy of a string theory
particle with momentum $p$, given by \bea {\cal
E}(p)=\sqrt{1+4g^2\sin^2\frac{p}{2}}\, . \eea The last term in the formula
(\ref{FSE}) can be understood as the finite-size
correction to the asymptotic, {\it i.e.} large $J$, dispersion
relation.

\smallskip

As in the general case, for a single excitation the value of $p$,
or equivalently the value of the rapidity $u(p)$
\bea\label{rapidity}
u(p)=\frac{1}{g}\cot\frac{p}{2}\sqrt{1+4g^2\sin^2\frac{p}{2}} \, ,
\eea is found from the exact Bethe equation $Y_{1*}(u)=-1$. As was
pointed out in section 2, when $J$ becomes large, the exact Bethe
equations  should turn into the asymptotic Bethe Ansatz equations,
which for a single excitation are simply \bea\label{ABA} e^{ipJ}=1
~~~~~~\Longrightarrow~~~~~~pJ=2\pi n\, ,~~~~~~ n=0,1,\ldots
\left[\frac{J}{2}\right]\, . \eea

Analogously, in various known integrable models the TBA approach
leads to the following expression for the total momentum
\begin{align}
P =\sum_{i=1}^Mp_i- \frac{1}{2\pi}\sum_{Q=1}^{\infty}\int {\rm
d}v\, \frac{d\tilde{\cal{E}}^Q}{dv}\log(1+Y_{Q}(v))\, ,
\end{align}
where here $\tilde{\cal{E}}^Q$ would be the energy of a mirror
$Q$-particle; the integral term in the large formula can be
viewed as a finite-size correction to the total momentum of string
theory particles. For a single magnon one gets
\begin{align}
\label{standardmomentum} P =p-
\frac{1}{2\pi}\sum_{Q=1}^{\infty}\int {\rm d}v\,
\frac{d\tilde{\cal{E}}^Q}{dv}\log(1+Y_{Q}(v))\, ,
\end{align}
where $p$ is determined by equations (\ref{ABA}). It might appear
slightly surprising that even for a single  particle its momentum
is modified by the finite-size correction. One might offer a
physical interpretation of $P$ as being the combined momentum of a
string particle and the mirror background. The correction term
would then be thought of as the back-reaction of the mirror
background to the particle; the former also gets into motion when a string theory
particle is excited.

\smallskip

On general grounds we could expect that in a finite-size theory
it is the momentum $P$ which must be quantized. Thus, we are led
to two apparently different quantization conditions
 \bea \label{Why} e^{iPJ}=1 ~~~~~~\Longleftrightarrow~~~~~~~Y_{1_*}(u)=-1
\eea which nevertheless should be compatible with each other. For
model I we will find perturbative evidence that this is indeed
the fact. However, in model II we will show that with the same
definition of $P$ only one of the two quantization conditions
holds, namely the second one.

\section{TBA equations for models I and II}
In this section we start by discussing the analytic properties of the
asymptotic solution and then show how these properties are taken into account in the excited state TBA equations. Finally, we summarize the complete set of excited state TBA equations for a
single magnon in both the twisted and untwisted cases. These
equations will be subsequently used to find NLO corrections to the
energy and momentum through the TBA approach.

\subsection{Analytic properties of the asymptotic solution}
The analytic properties of the asymptotic Y-functions play an
essential role in constructing the excited state TBA equations.
This is because the integration contours for the excited state TBA
equations are not known a priori. Demanding the TBA equations to
be compatible with the asymptotic solution provides the necessary
information to fix the integration contours\footnote{Of course,
this approach is based on the assumption that the exact solution develops no new singularities
in addition to those which are already present in the asymptotic solution.}.

\smallskip
In general we could expect that with the level-matching
condition relaxed, the analytic properties of the asymptotic solution
would become more intricate, and as it turns out this is indeed the case for the
models of interest here. In addition we will see that the analytic properties
of the corresponding solutions in models I and II are considerably
different. Now as the TBA equations essentially involve three types of
terms
\begin{equation}
\log{Y}, \qquad \log{(1 \pm Y)} \quad \mbox{and} \quad
\log{(1-Y^{-1})},
\end{equation}
this means that as far as integration along a contour in the
complex plane is concerned, we are interested in those
points where
\begin{equation}
Y(r) = 0, \infty ,\pm 1.
\end{equation}
When changing either the coupling constant $g$ or the momentum $p$
of a state these points $r$ move in the complex plane, and as we will
see, these points can either always lie on one side of the
integration contour, or cross it dynamically when changing either
$g$ or $p$. Because the next sections of the paper will be directly
concerned with the small $g$ behaviour of the TBA equations, we
relegate a discussion of so-called critical behaviour in $g$ to
appendix B, focusing here on the points $r$ that are relevant
already at weak coupling.

\smallskip

Before moving on to discussing the particular zeros and poles in
either model, we want to briefly comment on how zeroes and poles
of the asymptotic Y-functions and points where they equal $\pm 1$
are related, and how to relate these to zeroes and poles of the
transfer matrices; we will illustrate this by using the
$Y_{M|vw}$-functions as a simple example. Analogous relationships
hold for other Y-functions as well.

\smallskip

In both model I and model II, because the denominator remains
finite, zeroes of $Y_{M|vw}$ are inherited from zeroes of the
transfer matrices used to construct them\footnote{The asymptotic Y-functions are constructed in terms of transfer matrices based on the underlying symmetry group of the model \cite{Kuniba:1993cn,Tsuboi};  in the context of the string sigma model the corresponding asymptotic Y-functions were presented in \cite{GKV09}.}. The presently relevant $Y$-functions are constructed in terms of the transfer matrices as follows;
\begin{align}
Y_{M|w} & = \frac{T_{1,M} T_{1,M+2}}{T_{2,M+1}} \, ,  \, \, \, \, \, Y_{-} = -\frac{T_{2,1}}{T_{1,2}} \, ,  \, \, \,\, \,  Y_{+} = - \frac{T_{2,3}T_{2,1}}{T_{1,2}T_{3,2}} \notag \, ,\\
Y_{M|vw}& = \frac{T_{M,1} T_{M+2,1}}{T_{M+1,2}}  = \frac{T_{M,1} T_{M+2,1}}{T_{M+1,1}^- T_{M+1,1}^+ - T_{M,1} T_{M+2,1}}.\label{eq:YmvwinT} 
\end{align}
Here the $\pm$ superscript denotes the usual shift of the argument by $\pm i/g$, and of course all transfer matrices are implicitly the $\alg{sl}(2)$ ones. From the last expression we immediately see that a zero $r$ of $T_{M,1}$ is also a zero of $Y_{M|vw}$, and it is clear that shifting this zero by $\pm i/g$ we get the value of the
rapidity where a lower Y-function becomes equal to minus one;
$Y_{M-1|vw}(r^\pm) = -1$, where $r^\pm \equiv r\pm i/g$.

\smallskip

\begin{figure}
\begin{center}
{ \setlength{\unitlength}{1mm}
\begin{picture}(60, 70)
\put(10, 52){\vector(1, 1){10}} \put(10, 52){\vector(1, -1){10}}
\put(0, 50){$Y_{1|(v)w}$} \put(10, 42){\vector(1, 1){10}} \put(10,
42){\vector(1, -1){10}} \put(0, 40){$Y_{2|(v)w}$} \put(10,
32){\vector(1, 1){10}} \put(10, 32){\vector(1, -1){10}} \put(0,
30){$Y_{3|(v)w}$} \put(10, 22){\vector(1, 1){10}} \put(10,
22){\vector(1, -1){10}} \put(0, 20){$Y_{4|(v)w}$} \put(10,
12){\vector(1, 1){10}} \put(10, 12){\vector(1, -1){10}} \put(21.5,
61){$r_0$} \put(21.5, 51){$r_1$} \put(21.5, 41){$r_2$} \put(21.5,
31){$r_3$} \put(21.5, 21){$r_4$} \put(36, 52){\vector(-1, 0){10}}
\put(36, 42){\vector(-1, 0){10}} \put(36, 32){\vector(-1, 0){10}}
\put(36, 22){\vector(-1, 0){10}} \put(41, 61){$\times$} \put(37,
51){$1 + Y_{1|(v)w}^\pm$} \put(37, 41){$1 + Y_{2|(v)w}^\pm$}
\put(37, 31){$1 + Y_{3|(v)w}^\pm$} \put(37, 21){$1 +
Y_{4|(v)w}^\pm$} \put(4,13){.} \put(4,11.5){.} \put(4,10){.}
\put(22,13){.} \put(22,11.5){.} \put(22,10){.} \put(43,13){.}
\put(43,11.5){.} \put(43,10){.}
\end{picture}
}
\end{center}
\caption{Zeroes of $Y_{M|(v)w} (+1)$ and their interrelations.
Note that $r_M$ \emph{here} stands for the full set of zeroes of
$T_{M+1}$.} \label{fig:Ymvwroots}
\end{figure}
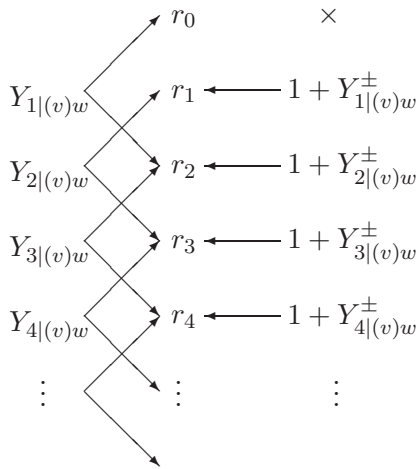

While the construction of $Y_{M|w}$ in terms of transfer matrices
does not immediately allow for a simple interpretation like the
one above, it should come as no surprise that similarly a zero of
$Y_{M|w}$ can give rise to points where $Y_{M-1|w}$ is equal to
$-1$. These relations for $Y_{M|w}$ and $Y_{M|vw}$ have been
summarized in the diagram in Figure \ref{fig:Ymvwroots}. Here we
will not be concerned with poles, as they play no role in either
model at small coupling.

\smallskip

With these ideas in mind, we are left with a rather elaborate discussion of all zeroes and poles of the asymptotic Y-functions in both models.

\subsubsection*{$\bullet$\   Model I}
In the untwisted case, the $Y_{M|vw}$-functions exhibit zeroes,
while $Y_+$ and $Y_-$ exhibit both zeroes and poles that have to
be taken into account already at weak coupling. As mentioned
earlier, these zeroes are directly linked to zeroes of the
transfer matrices; here each transfer matrix $T_M$ has a single
zero, which we will denote by $r_{M-1}$. These zeroes have a
natural ordering as $r_{M}
> r_{M+1}$ and they give rise to two zeroes of $Y_{M|vw}$, $r_{M-1}$
and $r_{M+1}$, and in addition to $Y_{M|vw}(r_M^\pm) = -1$. Also
$r_1$ is a zero of both $Y_+$ and $Y_-$, as also the transfer
matrices in their denominators remain finite at this point.
Finally, at $u^{\pm}$ the numerators and denominators of $Y_\pm$
have poles. However in $Y_-$ they cancel, whereas in $Y_+$ the
numerator has a pole of higher order, giving rise to poles of
$Y_+$ at $u^\pm$.

\smallskip

Furthermore $Y_+=1$ at the points $\mathcallig{r}_{0}$ and its conjugate
$\bar{\mathcallig{r}}_{0}$. These points give rise to a string of
poles of $Y_{M|w}$ and $Y_{M|vw}$, which are located at
$\mathcallig{r}_M$ and $\bar{\mathcallig{r}}_M$,
 where $\mathcallig{r}_M = \mathcallig{r}_0 + i M /g$.
These point play no role at weak coupling however; they are
discussed in more detail in appendix B.

\subsubsection*{$\bullet$\   Model II}
As mentioned above, twisting significantly modifies the analytic
properties of the asymptotic solution. First of all both $Y_{M|w}$
and $Y_{M|vw}$ now have zeroes. In the case of $Y_{M|vw}$,
according to our general discussion, these zeroes originate from
the ones of the transfer matrices; this time for $M>1$ each
transfer matrix has two zeroes, which are real or purely
imaginary, and come with opposite signs. Explicitly, we have
\begin{equation}
T_{M,1}(\pm r_{M-1}) = 0,
\end{equation}
where for $M=1$ we only have $T_{1,1}(- r_0)=0$. In fact, we were able
to guess the exact analytic form of these zeroes. It looks very
intriguing
\begin{equation}
\label{eq:rvw} r_M = \sqrt{u^2 - M(M+2) \mathcal{E} ^2/g^2},
\end{equation}
where $u$ is the Bethe root and $\mathcal{E}$ is nothing but the
asymptotic energy (\ref{FSE}); note that $r_0 = u$. For $p>0$, the
last formula can be rewritten as \bea r_M=\frac{{\cal
E}(p)}{g\sin\frac{p}{2}}\sqrt{1-(M+1)^2\sin^2\frac{p}{2}}\, . \eea
In general we thus have $Y_{M|vw}(\pm r_{M\pm1})=0$. For $M=1$, while the
story is a little different, the conclusion remains the same;
$Y_{1|vw}$ still has $u$ as a zero although it is not a zero of
$T_{1}$, this is due to the pole of $T_2$ at $u^-$. Of course,
these zeroes give rise to points $r_M^{\pm}$ where $Y_{M|vw}(\pm
r_M^\pm) = -1$.

\smallskip

Moving on, all $Y_{M|w}$-functions turn out to have two zeroes
which are always real. We will denote these zeroes by $\rho_{M-1}$
and $\rho_{M+1}$;
\begin{equation}
Y_{M|w}(\rho_{M-1})=Y_{M|w}(\rho_{M+1})=0.
\end{equation}
Again we were able to guess their form, which fascinatingly enough
is given by
\begin{equation}
\rho_M = - u_{M+1}
\end{equation}
where
\begin{equation}
u_{M} \equiv  \tfrac{M}{g} \cot{\tfrac{M p}{2}}
\sqrt{1+4 (\tfrac{g}{M})^2 \sin^2{\tfrac{M p}{2}}},
\end{equation}
where one immediately recognizes that $\rho_0 = -u$. Similar to
$Y_{M|vw}$, $1 + Y_{M|w}$ has zeroes at the points $\rho_M \pm
i/g$.

\smallskip

\begin{figure}[h]
\begin{center}
\includegraphics[width=5.8in]{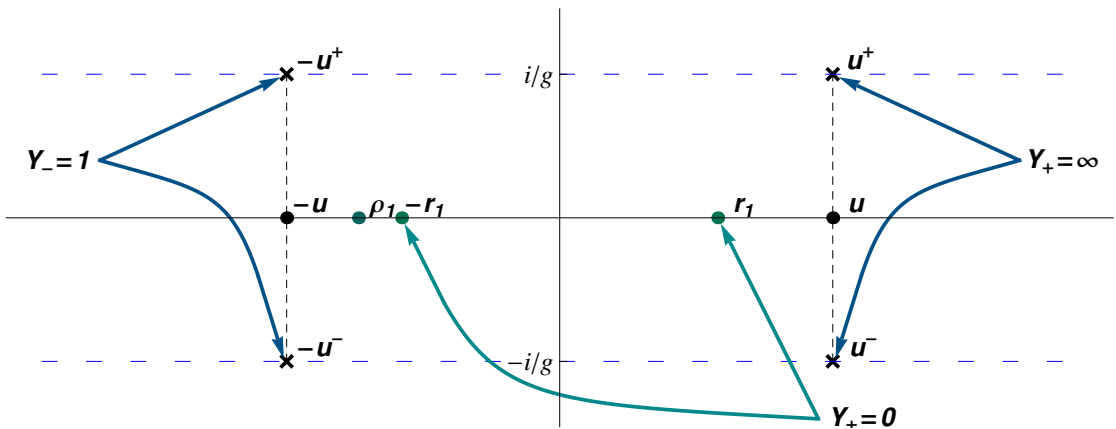}
\end{center}
\caption{Special points for $Y_\pm$ in the complex $v$ plane. Note
that $r_1$ can also be imaginary as opposed to real as is
illustrated here.} \label{fig:Ypmroots}
\end{figure}

Finally the $Y_\pm$ functions each have the same two
zeroes\footnote{Not surprising since $Y_+$ and $Y_-$ share a
common factor, which then of course is the source of the zeroes.},
$Y_+(\pm r_1)=Y_-(\pm r_1)=0$. The points $\pm r_M^\pm$ which gave
rise to roots of $1 + Y_{M|vw}$, here for $M=0$ split nicely into
zeroes of $1-Y_-$ at $-u^\pm$ and poles of $Y_+$ at $u^\pm$ giving
a very symmetric situation, as illustrated in Figure
\ref{fig:Ypmroots}. In addition both $Y_+$ and $Y_-$ have a pole at $\rho_1$.

For the reader's convenience, an overview of the relevant roots
and poles in both models is presented in Table
\ref{tab:rootsandpoles}.

\begin{table}
\begin{center}
\begin{tabular}{|c|c|c|c|c|c|}
\hline Y-function  & Zeros I    & Poles I &  Zeros II  & Poles II & Cuts\\
\hline
\hline $Y_+$       & $r_1$        &  $u^{\pm}$             & $\pm r_1$  & $u^{\pm}, \rho_1$ & $0,\pm 2$\\
\hline $Y_+ - 1$     & $\mathcallig{r}_{0}\, ,\bar{\mathcallig{r}}_0$        &             & - & & \\
\hline $Y_-$       & $r_1$        &        -       & $\pm r_1$ & $\rho_1$ & $0,\pm 2$\\
\hline $Y_- - 1$     & -        &             & $-u^\pm$ & & \\
\hline $Y_{M|w}$   &        -      &   $ \mathcallig{r}_M\, ,\bar{\mathcallig{r}}_M$         & $\rho_{M\pm1}$ & - & $\pm M,\pm (M+2)$\\
\hline $Y_{M|w} +1$& - &            & $\rho_{M}^\pm$ & & \\
\hline $Y_{M|vw}$  &  $r_{M\pm 1}$&   $ \mathcallig{r}_M\, ,\bar{\mathcallig{r}}_M$           &  $\pm r_{M\pm 1}$ & - & $\pm M,\pm (M+2)$\\
\hline $Y_{M|vw} +1$  &  $r_{M}^{\pm}$ &            & $\pm r_{M}^{\pm}$ & & \\
\hline
\end{tabular}
\end{center}
\caption{Essential analytic structure of the asymptotic solution
for a single magnon in both the twisted (right) and untwisted
(left) case. A dash indicates absence of a zero or pole, while the
empty spots would have information that is already in the table.
The branch cuts are $(-\infty,2] \cup[2,\infty)$, shifted up and
down in the complex plane by the multiples of $i/g$ indicated.
N.B. $-u^\pm$ should be read as $-u\pm \frac{i}{g}$ everywhere.}
\label{tab:rootsandpoles}
\end{table}

\subsection{The simplified TBA equations}

In order to obtain the excited state TBA equations, following
\cite{Arutyunov:2009ax}, we employ the contour deformation trick.
In short we assume that the ground state and excited state TBA
equations differ only by the choice of the integration contours,
and upon deforming the integration contours of the excited state
TBA equations to the ground state ones, we pick up contributions
of extra singularities, leading to the appearance of new driving
terms in the excited state TBA equations. This procedure is very
similar in spirit to the Dorey-Tateo trick \cite{DT96}. Below we
discuss the choice of the integration contours in detail.

\smallskip

\begin{figure}[h]
\begin{center}
\includegraphics[width= 6in]{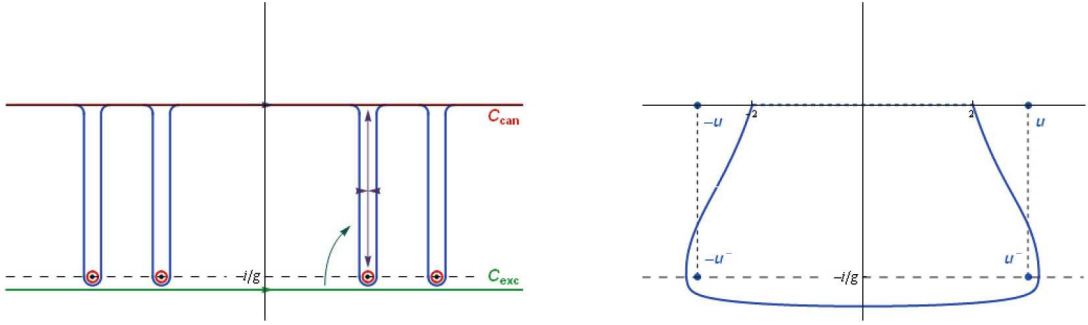}
\end{center}
\caption{The integration contour for all Y-functions except
$Y_\pm$ (left) and the integration contour for $Y_\pm$ (right).}
\label{fig:contours}
\end{figure}

The integration contours for all functions $Y_Q$, $Y_{M|w}$ and
$Y_{M|vw}$ can be taken the same, hence we will speak of
\emph{the} integration contour for $Y_Q$, $Y_{M|w}$ and $Y_{M|vw}$
functions. This contour needs to be chosen such that it
encompasses the pole of $\log{1+Y_Q}$ at the Bethe roots in the string region, which is naturally
motivated by the fact that this gives a driving term for each Bethe root, see \cite{Arutyunov:2009ax} for an explicit discussion. With this requirement it is only
natural that the integration contour should run a little bit below
the line $-i/g$ after deforming it, and in fact we find that accepting or discarding
contributions of zeroes and poles according to this choice of the
contour, we obtain a set of TBA equations which is perfectly
satisfied by the asymptotic solution. Returning the integration
contour to the real mirror line, \emph{we pick up a
contribution from all zeroes and poles that lie between the real line and
the line $-i/g$, as well as the usual Bethe root contributions}. The integration contour and the extra
contributions that arise by returning the integration contour for
excited states, $C_{exc}$, to the real line, $C_{can}$, are
illustrated in the left panel of Figure \ref{fig:contours}.

\smallskip

Analogously, for $Y_+$ and $Y_-$ the integration contour needs to
be picked essentially in the same manner, with of course the only
exception that it should start at $-2$ and end at $2$. In
particular, it is chosen in such a way that we pick up both the
pole of $Y_+$ at $u^-$ and the zero of $1 - Y_-$ at $-u^-$, as
illustrated in the right panel of Figure \ref{fig:contours}.

\smallskip

Now we would like to briefly come back to the zeroes of $Y_{M|vw}$ in model
II. Looking at equation (\ref{eq:rvw}), we see that they are either real
or imaginary depending on a state ($p$) and the number $M$ under
consideration. Also, considered as a function of momentum, a given
zero will go from real to imaginary values as the momentum is
increased ($0\leq p\leq \pi$). When it is real or imaginary but
lies between $-i/g$ and the real axis, we need to take its
contribution into account in the TBA equations. However, from
equation (\ref{eq:rvw}) we see that as momentum is increased, a given
zero will move out of the physical strip $-1/g\leq {\rm Im}\leq
1/g$ and eventually no longer needs to be taken into account.
\emph{So the appearance of the zeroes $r_M$ in the TBA is
state-dependent.} Of course, the zeroes also move as $g$ changes, but for now we are concerned only with the small $g$ behaviour just discussed; we refer the reader to appendix B
for further details on the behaviour in $g$.

\smallskip

We would like to note that the kernels and S-matrices appearing
below have been defined and are completely listed in
\cite{Arutyunov:2009ax}; for the sake of brevity we refrain from
reproducing the list here and hence refer the reader there for
their exact form.

\smallskip

Taking the integration contours as indicated above together with
the simplified TBA equations for the ground state derived in
\cite{AF09b,AF09d}, and returning the integration contours to the
real line of the mirror theory, we obtain the excited state TBA
equations for a single magnon. For both comparison purposes and
brevity, we will discuss model I and model II side by side,
splitting the TBA into the always present ground state and $Y_1$
contributions, and the extra terms that come from zeroes and poles
discussed above which differ between the twisted and untwisted
case, see Table \ref{tab:rootsandpoles}. Please note that in what
follows for model II, for completeness we are always writing down the
contributions from zeroes $r_M$; of course they are not present if
the corresponding zero falls outside the lower-half of the
physical strip.
We then find the following sets of integral equations.\vspace{10pt}\\
\bigskip
 \noindent
$\bullet$\ $M|w$-strings; $\ M\ge 1\ $, $Y_{0|w}=0$
\begin{equation*}
V_{M|w} \equiv \log Y_{M|w} -  \Blueuline{\log(1 +  Y_{M-1|w})(1 +
Y_{M+1|w})\star s}
 - \delta_{M1}\, \log{1-{1\ov Y_-}\ov 1-{1\ov Y_+} }\hstar s \,,~~~~~
\end{equation*}
where for the two different models we have
\begin{equation}
\label{Yforws} V_{M|w}=
\begin{cases}
0 & \mbox{I} \\
- \Blueuline{\log{S(\rho_{M-1}^-)S(\rho_{M+1}^-)}}& \mbox{II}
\end{cases}
\end{equation}
The extra contributions in model II arise from the zeroes
of $1 +  Y_{M-1|w}$ and $1 +  Y_{M+1|w}$,
as indicated by the blue underlining; color coding has been used
in all equations below for the readers convenience. Therefore, we see that
while for model I these equations coincide with the ground state ones,
for model II \emph{they do not}.\vspace{10pt}\\
\bigskip
 \noindent
$\bullet$\ $M|vw$-strings; $\ M\ge 1\ $, $Y_{0|vw}=0$ \bea
\nonumber \hspace{-0.3cm}V_{M|vw} \equiv &&\log Y_{M|vw}(v)
~~~~~\\\nonumber &&+ \log(1 +  Y_{M+1})\star s -
\RoyalBlueuline{\log(1 +  Y_{M-1|vw} )(1 +  Y_{M+1|vw})\star
s}~~~~~\\\nonumber &&- \Greenuline{\delta_{M1}  \log{1-Y_-\ov
1-Y_+}\hstar s}\, ,~~~~~\\\nonumber \eea where for the two
different models we have
\begin{equation}
\label{Yforvw} V_{M|vw} =
\begin{cases}
- \Greenuline{\delta_{M1} \log{S(u^- - v)}} -  \RoyalBlueuline{\log{S(r_{M-1}^- - v)S(r_{M+1}^- -v)}} & \mbox{I} \\
- \Greenuline{\delta_{M1} \log{S(\pm u^- - v)}} -
\RoyalBlueuline{\log{S(\pm r_{M-1}^- - v)S(\pm r_{M+1}^- -v)}} &
\mbox{II}
\end{cases}
\end{equation}
Here and in what follows below we \emph{implicitly assume a sum} over the $\pm$ sign in front of the zeroes.
Given the shape of the integration contour, for $M=1$ the first term in both models comes from the pole of $Y_+$
at $v=u^-$, and the zero of $1-Y_-$ at $v=-u^-$ in model II, whereas the second terms come from the zeroes
of $1 +  Y_{M-1|vw}$ and $1 +  Y_{M-1|vw}$, {\it cf.} Table \ref{tab:rootsandpoles}.
So we see that also at weak coupling, single particle states
have additional contributions in the TBA equations for $vw$-strings. Finally,
we see that the pole of $S(\pm u^- - v)$ cancels the  zero of $Y_{1|vw}(v)$ at $v = \pm u$ as it must,
and that  equation (\ref{Yforvw}) is compatible with the reality condition for Y-functions.\vspace{10pt}\\
\bigskip
 \noindent
$\bullet$\   $y$-particles \bea \nonumber
V_\pm \equiv && \log {Y_+\ov Y_-}(v) +\log S_{1_*y}(u,v)  -\log(1 +  Y_{Q})\star K_{Qy} = 0\,,~~~~~~~ \\
\nonumber
V_\mp \equiv && \log {Y_- Y_+}(v) + \log {\big(S_{xv}^{1_*1}\big)^2\ov S_2}\star s(u,v) -\Aquamarineuline{2\log{1 +  Y_{1|vw} \ov 1 +  Y_{1|w} }\star s} \\
\nonumber &&+ \log\left(1+Y_Q \right)\star K_Q+ 2 \log(1 +
Y_{Q})\star K_{xv}^{Q1} \star s\,,~~~~ \eea where for the two
different models we have
\begin{equation}
\la{Yfory} V_{\mp} =
\begin{cases}
-\Aquamarineuline{2 \log{S(r_1^- - v)}}& \mbox{I} \\
-\Aquamarineuline{2 \log{S(\pm r_1^- - v)} - 2 \log{S(\rho_1^- -
v)}}& \mbox{II}
\end{cases}
\end{equation}
Here the additional contributions come from the term $\log{1 +
Y_{1|vw} \ov 1 +  Y_{1|w} }$ and we have used the following
notation
 \bea\nonumber
 &&\log  {\big(S_{xv}^{1_*1}\big)^2\ov S_2}\star s(u,v) \equiv  \int_{-\infty}^\infty\, dt\, \log  {S_{xv}^{1_*1}(u,t)^2\ov S_2(u-t)}\, s(t-v)
 \,.~~~~~
\eea
Moreover, $S_{1_*y}(u_{j},v) \equiv S_{1y}(z_{*j},v)$ is a shorthand notation
for the S-matrix with the first and second arguments in the string and mirror regions, respectively.
The same convention is used for other S-matrices and kernels.\vspace{10pt}\\
\bigskip
 \noindent
$\bullet$\ $Q$-particles

The TBA equations for $Q$-particles can be cast in particularly
convenient form called the hybrid equations; their general form has been derived in \cite{Arutyunov:2009ax}. It is here that we find another significant difference from level-matched
cases studied previously, as well as between model I and II. We
find that in order for the asymptotic Y-functions to satisfy the
following integral equations for one-particle states, for model I
the length parameter $L_{\scriptscriptstyle TBA}$ has to be equal
to the angular momentum\footnote{In other words, the same as for
the ground state \cite{FS}.} $J$, whereas in model II we find that
$L_{\scriptscriptstyle TBA}=J+1$.

\smallskip

This merits a further discussion, as this behaviour is of course
far from an accident. We have briefly investigated the hybrid
equations for two and three level-unmatched magnons in the
untwisted theory, and there we consistently find the relation to
be $L_{\scriptscriptstyle TBA}=J$. On the other hand, for
level-matched states it was observed in \cite{Arutyunov:2009ax}
that \mbox{$L_{\scriptscriptstyle TBA}=J+2$.} Thus, in the
untwisted case there is a clear difference between physical
(level-matched) and unphysical (level-unmatched) states, rather
than a difference between different numbers of magnons. Doing the
twist removes some of the differences between the level-matched
and unmatched cases, hence leaving us in some sense somewhere in
between\footnote{We will see more examples of how the twist brings
us `closer' to the level-matched case later.} with
$L_{\scriptscriptstyle TBA}=J+1$. In the equations below we thus
have
\begin{equation*}
L_{\scriptscriptstyle TBA}=\begin{cases}J & \mbox{I}~~~\mbox{level-unmatched}\\
J+1 & \mbox{II} \\
J+2 & \mbox{I}~~~\mbox{level-matched}
\end{cases}
\end{equation*}\\
The \emph{hybrid equations for $Q$-particles} are now given by
\begin{align}
V_Q \equiv &\log Y_Q(v)+ L_{\scriptscriptstyle TBA}\, \tH_{Q} - \log \left(1+Y_{Q'} \right) \star \(K_{\sl(2)}^{Q'Q} - 2 \, s \star K^{Q'-1,Q}_{vwx} \)  + \log S_{\sl(2)}^{1_*Q}(u,v) \notag\\
&  - \Blueuline{ 2 \log \(1 + Y_{1|vw}\) \star s \hstar K_{yQ}} - \RoyalBlueuline{ 2 \, \log \(1 + Y_{Q-1|vw}\) \star  s }  \label{TbaQsl2H} \\
&  + \Aquamarineuline{2  \log{1-Y_-\ov 1-Y_+} \hstar s \star
K^{1Q}_{vwx}}  - \Greenuline{ \log {1- \frac{1}{Y_-} \ov 1-
\frac{1}{Y_+} } \hstar K_{Q} }-  \Greenuline{ \log \big(1-
\frac{1}{Y_-}\big)\big( 1 - \frac{1}{Y_+} \big) \hstar_{p.v.}
K_{yQ} }\,, \notag
\end{align}
where for model I we have
\begin{align}
V_Q = & -\Blueuline{2 \log S\hstar_{p.v.} K_{yQ} (r_1^-,v)} -  \RoyalBlueuline{ 2 \log S (r_{Q-1}^-,v) } + \\
 &  \quad +  \Aquamarineuline{2 \log S\star_{p.v.} K^{1Q}_{vwx}(u^-,v) }-  \Aquamarineuline{\log S^{1Q}_{vwx}(u,v)}\nonumber \,,
\end{align}
while for model II we have
\begin{align}
V_Q = & -\Blueuline{2 \log S\hstar_{p.v.} K_{yQ} (\pm r_1^-,v)} - \RoyalBlueuline{ 2 \log S (\pm r_{Q-1}^-,v)} + \\
 & \quad +   \Aquamarineuline{ 2 \log S\star_{p.v.} K^{1Q}_{vwx}(\pm u^-,v) }\pm \Aquamarineuline{\log S^{1Q}_{vwx}(\pm u,v)} + \nonumber \\
 & \quad \quad -\Greenuline{  \log{S_{Q}}(-u^-,v) }- \Greenuline{ \log{S_{yQ}}(-u^-,v)}\nonumber \,.
\end{align}

\begin{figure}[h]
\begin{center}
\includegraphics[width=4in]{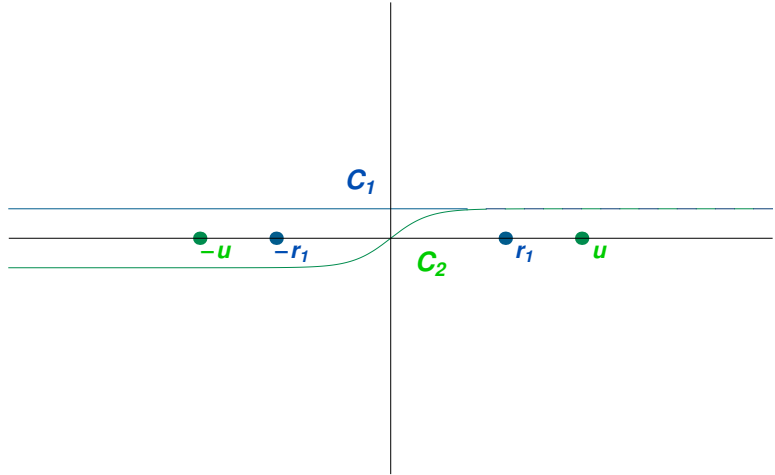}
\end{center}
\caption{The integration contours for the kernels $\log S\star
K_{yQ}$ and $\log S\star K^{1Q}_{vwx}$ in the hybrid equations.}
\label{fig:pmiepscontour}
\end{figure}

\noindent In the above $K^{0,Q}_{vwx}=0$ and $Y_{0|vw}=0$, which
implies that the $\log S (\pm r_{0}^-,v)$ terms are also not
present. The principal value prescriptions are required due to the
pole of $S(v)$ at $v=-i/g$, and are taken in accord with the
integration contour $C_1$ in the term $\log S\star K_{yQ}$ running
slightly above the real line, while for $ \log S\star
K^{1Q}_{vwx}$ the contour $C_2$ runs slightly above the real line
in model I (like $C_1$) and slightly below and above the negative
and positive half real axis in model II respectively,  as
illustrated in Figure \ref{fig:pmiepscontour}. We would like to
note that the principal value prescription for the terms involving
$K_{yQ}$ give extra contributions, but these precisely cancel
against one another. All extra terms above follow from the same
reasoning as for the other TBA equations; once more we have explicitly indicated their origin via color coding.

\subsection{Exact Bethe equations}
Analytically continuing the hybrid equations in the rapidity $v$
to the string region needs to be carefully done as certain terms
in the equations below have branch cuts when considered as a
function of $v$. This continuation has been carried out in
\cite{Arutyunov:2009ax} for the Konishi state, and here we follow
the same logic to obtain
\begin{align}
&V_{Q_*} \equiv \pi i(2n+1) - i L_{\scriptscriptstyle TBA}\, p - \log \left(1+Y_{Q} \right) \star \(K_{\sl(2)}^{Q1_*} - 2 \, s \star K^{Q-1,1_*}_{vwx} \) + \, \log S_{\sl(2)}^{1_*1_*}(u,u)\nonumber\\
&\quad - \Blueuline{2 \log \(1 + Y_{1|vw}\) \star \( s \hstar
K_{y1_*} + \ts\)} + \Greenuline{2  \log{1-Y_-\ov 1-Y_+} \hstar s
\star_{p.v.} K^{11_*}_{vwx}} +  \Greenuline{2  \log{1-Y_-\ov
1-Y_+} \hstar s}
\notag \\
&\quad - \RoyalBlueuline{\log {1- \frac{1}{Y_-} \ov 1-
\frac{1}{Y_+} } \hstar K_{1}} -  \Aquamarineuline{\log \big(1-
\frac{1}{Y_-}\big)\big( 1- \frac{1}{Y_+} \big) \hstar
K_{y1_*}}\label{Tba1sl2B}
\end{align}
where for model I we have
\begin{align}
V_{Q_*} = &  - \Blueuline{2 \log S \hstar_{p.v.} K_{y1_*} (r_1^-,u)} - \Blueuline{2 \log S (r_1^- -i/g- u)} \label{eq:EBmodelI}\\
& \quad + \Greenuline{2 \, \log {\rm Res}(S)\star K^{11_*}_{vwx}
(u^-,u)} - \Greenuline{2 \log{{2i\ov g}\, {x^--{1\ov x^-}\ov x^+ -
x^-}}}\nonumber\,,
\end{align}
and for model II we have
\begin{align}
V_{Q_*} = &  - \Blueuline{2 \log S \hstar_{p.v.} K_{y1_*} (\pm r_1^-,u)}-\Blueuline{2 \log S (\pm r_1^-  -i/g - u)} +  \label{eq:EBmodelII}\\
& \quad + \Greenuline{2 \, \log {\rm Res}(S)\star K^{11_*}_{vwx}
(u^-,u)} - \Greenuline{2 \log{{2i\ov g}\,
{x^--{1\ov x^-}\ov x^+ - x^-}}} \nonumber \\
& \quad \quad + \Greenuline{2 \, \log S\star_{p.v.} K^{11_*}_{vwx} (-u^-,u)} + \Greenuline{\log{S^{11_*}_{vwx}}(-u,u)} + \Greenuline{\log{S(-u^-,u)}} + \nonumber \\
& \quad \quad \quad  -  \RoyalBlueuline{\log{S_{1}}(-u^-,u)}  -
\Aquamarineuline{\log{S_{y1_*}}(-u^-,u)} \nonumber\,.
\end{align}
In the above we use the notations \bea
&&\log {\rm Res}(S)\star K^{11_*}_{vwx} (u^-,v) = \int_{-\infty}^{+\infty}{\rm d}t\,\log\Big[S(u^- -t)(t-u)\Big] K_{vwx}^{11*}(t,v)\,,~~~\notag\\
&&\ts(u)=s(u^-)\notag\,, \eea the momentum of the magnon is $p = i
\tH_{Q}(z_{*})=-i\log{x_s(u+{i\ov g})\ov x_s(u-{i\ov g})}$, and
the second argument in all kernels in \eqref{Tba1sl2B} is the
Bethe root $u$.

\smallskip

For a single magnon, the exact Bethe equation in the $\sl(2)$
sector can be written in the form
 \bea\nonumber
\pi i(2n+1)=i J\, p - \log S_{\sl(2)}^{1_*1_*}(u,u)+i{\cal R
}\la{BYsl2single}
 \,,~~~~~~
\eea where ${\cal R}$ encodes a correction to the asymptotic Bethe
equation represented by the remaining terms. For asymptotic
solution we should find ${\cal R}^{\rm asym}=0$, which also serves
as a non-trivial check of the correctness of the analytic
continuation procedure.

\smallskip

Discarding  for large $J$ the exponentially small $Y_Q$-functions
in the right hand side of equation (\ref{Tba1sl2B}) and subtracting from
equation (\ref{Tba1sl2B}) the contribution of the asymptotic Bethe
equation, we find ${\cal R}^{\rm asym}$ for models I and
II. We recall that the difference between $L_{\scriptscriptstyle TBA}$
and $J$ that enters ${\cal R}^{\rm asym}$ differs between the models; $L_{\scriptscriptstyle TBA}-J=0$ for
model I and $L_{\scriptscriptstyle TBA}-J=1$ for model II.
Substituting the asymptotic Y-functions into the respective ${\cal
R}^{\rm asym}$, we find that the corresponding quantity perfectly
vanishes for both models.

\section{Finite-size corrections to the dispersion: Model I}

For a fixed $J$ and small $g$, the $Y_Q$-functions are small. The
finite-size correction term then provides the so-called wrapping
correction to the energy, or, equivalently, to the scaling
dimension of a gauge theory operator at weak coupling. The goal of
the next two sections is to compute the leading wrapping
corrections to the asymptotic energy and momentum of a single
magnon for models I and II.

\smallskip

When expanding energy and momentum for small $g$ around the
asymptotic solutions $Y^{o}_Q$, we obtain the leading order
corrections to the energy and momentum {\small
\begin{align}\label{eqn;Luscher} E_{\rm LO} =
-\frac{1}{2\pi}\sum_{Q=1}^{\infty}\int {\rm d}v\,  Y^{o}_{Q}(v)\,
, ~~~~ P_{\rm LO} = -\frac{1}{2\pi}\sum_{Q=1}^{\infty}\int {\rm
d}v \frac{2v}{Q^2+v^2}Y_{Q}^{o}(v)\, .
\end{align}
}We recall that formula (\ref{eqn;Luscher}) for the energy
follows from L\"uscher's perturbative approach and it was shown to
successfully describe the four-loop corrections to the anomalous
dimension for twist two operators \cite{BJ08,Bajnok:2008qj}. When
the level-matching condition is imposed, the functions $Y_{Q}^{o}(v)$
are symmetric with respect to $v\to -v$ and therefore, $P_{\rm
LO}$ vanishes, as it should. For a single magnon however, the
$Y_{Q}^{o}(v)$ are no longer symmetric which leads to
non-trivial values of $P_{\rm LO}$.

\subsection{LO wrapping correction to the energy and momentum}
In model I we deal with periodic boundary conditions for all
fields (no twist). Using the untwisted transfer matrices
$T_{Q,1}$, we find that in the limit $g\to 0$ the $Y_Q$-functions
associated to a single magnon with rapidity $u$ are
\begin{align} Y^{o}_Q(v) = \frac{g^{2J}}{(v^2 +
Q^2)^J} \frac{4 Q^2 \left(u-\sqrt{u^2+1}\right)^2
(v-u)^2}{[(Q-1)^2+(v-u)^2][(Q+1)^2+(v-u)^2]}\, .
\end{align}
The energy integral (\ref{eqn;Luscher}) can be computed by
residues for any fixed value of $J$ and the first wrapping
correction to the asymptotic dispersion relation is then found by
taking the sum over the bound state number $Q$.

\smallskip

In appendix C we present the full result of the leading order wrapping correction depending on the rapidity $u$. Here we present only the result upon substituting the solution of the Bethe equations $e^{ipJ}=1$. We find for model I the following compact result for $J>4$
\begin{align}\label{eqn;ExactWrappingEonshell}
E_{\rm LO}^{\rm I}(J) =&~ g^{2J}\,\sum_{m=1}^{\left\lfloor
\frac{J}{2}\right\rfloor+1}C_m(J) \zeta (2 J-2 m-1)\, .
\end{align}
Here the coefficients $C_m(J)$, written in terms of particle
momentum, take the form
\begin{align}
 C_m(J)=&\frac{(-1)^{m+1} \Gamma(2J-2m)}{2^{2J-2m-1}\Gamma(J)\Gamma
(J-2m+2)}\tan^2\frac{p}{4}\Big(\sin\frac{p}{2}\Big)^{2m-1}\times \\
\nonumber &\hspace{2cm} \times\left[(m-1) \sin\frac{p}{2}\cos
mp-(J-m)\cos\frac{p}{2}\sin mp\right].
\end{align}
The above formula does not hold for $J\leq 4$ because of
divergent terms. However, for these values of $J$ the wrapping
corrections are easily found to be
\begin{align}
E_{\rm LO}^{\rm I}(4)&=2 \zeta (3) \sin ^6\frac{p}{4} \cos^2\frac{p}{4}(2 + 4 \cos p+\cos 2 p)-\frac{5}{4} \zeta (5) \sin ^4 \frac{p}{4} (1 + \cos p)\\
E_{\rm LO}^{\rm I}(3)&=-\frac{3}{16} \zeta (3) \sin ^2p \tan ^2\frac{p}{4}\\
E_{\rm LO}^{\rm I}(2)&=\sin ^4\frac{p}{4}\cos p.
\end{align}
It is worth stressing that for
generic $p$ the integral for $E_{\rm LO}^{\rm I}(J)$ also gives
rise to polygamma functions. However, these cancel out on
solutions of the Bethe equation leaving only a sum of
$\zeta$-functions.

\smallskip

We conclude this section by mentioning one important subtlety
concerning the level-matching condition. As we have seen from the
single magnon example and as can be verified for the case of
several particles, without level matching imposed the wrapping
correction in model I expands at weak coupling as \bea E^{\rm I
}_{\rm LO}=g^{2J}E^{(2J)}+\ldots
\,=g^{2L_{TBA}}E^{(2L_{\scriptscriptstyle TBA})}+\ldots , \eea
since, as we have shown in section 3, in the level-unmatched case
$L_{\scriptscriptstyle TBA}=J$. On the other hand, with the level
matching imposed, one finds that the leading term of order
$g^{2J}$ cancels out\footnote{The transfer matrix $T_{Q,1}$ is
constant as $g\to 0$ in the level-unmatched case and behaves as
$g^2$ when the level matching is imposed.} and the expansion of
the wrapping corrections starts from $g^{2J+4}$.\ However, as we
have seen, for the level-matched case the TBA equations require to
set up $L_{\scriptscriptstyle TBA}=J+2$. Thus, \bea E_{\rm
LO}^{\rm I }=g^{2J+4}E^{(2J+4)}+\ldots
\,=g^{2L_{TBA}}E^{(2L_{TBA})}+\ldots , \eea {\it i.e.} in both
cases the expansion starts from $g^{2L_{\scriptscriptstyle TBA}}$.

\smallskip

Now we present the leading finite size correction to the momentum.
To make the relation to $E_{\rm LO}$ more transparent, we
introduce
\begin{align}
&E_{\rm LO}^{\rm I} = \frac{(u-\sqrt{u^2+1})^2}{2\pi}\mathscr{E},&&
P_{\rm LO}^{\rm I} = \frac{(u-\sqrt{u^2+1})^2}{2\pi}\mathscr{P}.
\end{align}
From the identity
\begin{align}
\label{IdentityI}
\partial_u \frac{Y_Q^0(v)}{(u-\sqrt{u^2+1})^2} + \frac{2Jv}{v^2+Q^2} \frac{Y_Q^0(v)}{(u-\sqrt{u^2+1})^2}
+ \frac{\partial_v  Y_Q^0(v)}{(u-\sqrt{u^2+1})^2} = 0,
\end{align}
it is then easily shown that
\begin{align}
\label{PE} J \mathscr{P} = -\,\partial_u \mathscr{E}.
\end{align}
It is now straightforward to find the exact expression for the
wrapping corrected momentum by differentiating the formula
(\ref{eqn;ExactWrappingE}). Finally, we point out that formula
(\ref{PE}) implies the following relation between the leading
corrections to the energy and momentum
\begin{align}
\label{PEwrap} J P^{\mathrm{I}}_{\rm LO} =
2\sin\frac{p}{2}\Big(\sin\frac{p}{2}\frac{\pa}{\pa p}E^{\mathrm{I}}_{\rm
LO}-E^{\mathrm{I}}_{\rm LO}\Big).
\end{align}
We would like to note that on solutions of the Bethe equations, the expression for $P^{\mathrm{I}}_{\rm LO}$ generically contains polygamma terms.

\subsection{Quantization of momentum and exact Bethe equation}
To test the equivalence of the two quantization conditions
(\ref{Why}), we now need to independently compute the leading
finite-size correction to the asymptotic Bethe equations. We will
do this analytically by applying the method developed in
\cite{Balog:2010vf,Balog:2010xa}. It is based on finding a
connection between $Y^{o}_Q$ and the transfer matrix of the
Heisenberg spin chain. For a single excitation with rapidity $u$
the transfer matrix of the Heisenberg model is
\begin{align}
t_M(v) = (M+1)(v-u).
\end{align}
It is readily checked that it satisfies the equation
\begin{align}
t_{M}(v+i)t_{M}(v-i) = t_{M+1}(v)t_{M-1}(v) +
t_{0}(v+(M+1)i)t_{0}(v-(M+1)i).
\end{align}
The corresponding Y-function is then defined as
\begin{align}
y_M(v) = \frac{t_{M+1}(v)t_{M-1}(v)}{t_{M}(v+i)t_{M}(v-i) -
t_{M+1}(v)t_{M-1}(v) } = \frac{M(M+2)(v-u)^2}{(M+1)^2+ (v-u)^2},
\end{align}
which, in turn, satisfies the functional equation
\begin{align}
y_M(v+i)y_M(v-i) = [1+y_{M+1}(v)][1+y_{M-1}(v)].
\end{align}
Obviously, each $y_{M}(v)$ has a double zero at $r_M = u$. Furthermore, $y_{M}(v)$ also enjoys the normalization condition
\begin{align}
y_M(u-i)=-1.
\end{align}
In terms of the quantities introduced above, the leading order of
$Y^{o}_Q(v)$ can be conveniently written as
\begin{align}\label{eqn;Y0inXXX1st}
Y^o_Q(v) = -\frac{4g^{2J}}{(v^2+Q^2)^{J}}(\sqrt{1+u^2}-u)^2j_Q,
\end{align}
with
\begin{align}
j_Q =
\frac{t_{Q-1}(v)^2}{t_0(v+(Q+1)i)t_0(v+(Q-1)i)t_0(v-(Q-1)i)t_0(v-(Q+1)i)}.
\end{align}
Moreover, expanding $Y_{M|vw}^o(v)$ at weak coupling, we find that
the leading term coincides with $y_M(v)$. This is all that is needed to repeat the computation of
\cite{Balog:2010vf,Balog:2010xa} for the case at hand.

\smallskip

Following the same strategy as employed in \cite{Arutyunov:2010gb}, we
employ the exact Bethe equations in their hybrid form
(\ref{Tba1sl2B}). These equations will allow us to find the
leading finite-size correction $\delta u$ to the asymptotic magnon
rapidity $u\equiv u(p)$, the latter given by formulae
(\ref{rapidity}) and (\ref{ABA}). Expanding (\ref{Tba1sl2B})
around its asymptotic solution, we get
\begin{align}
2\pi n = J p  + \delta \mathcal{R} +\mathcal{O}(g^{2J+1}).
\end{align}
Here $\delta {\mathcal R}$ is a leading correction of order
$g^{2J}$ to the asymptotic value of  $\mathcal{R}$, the latter
being identically zero. As in the two-particle case
\cite{Arutyunov:2010gb}, the term $\delta \mathcal{R}$ is
naturally split into three parts
\begin{align}
\delta \mathcal{R}= \delta \mathcal{R}^{(1)} + \delta
\mathcal{R}^{(2)} + \delta \mathcal{R}^{(3)}\, ,
\end{align}
where
\begin{align}
& \delta \mathcal{R}^{(1)}  =  \frac{1}{\pi} \sum_{Q=1}^{\infty} \int {\rm d}v\, Y^o_Q(v)\frac{v-u}{(Q+1)^2+(v-u)^2},\\
& \delta \mathcal{R}^{(2)} =  \frac{1}{\pi} \sum_{Q=1}^{\infty}
\int {\rm d}v\,
Y^o_{Q+1}(v)\left\{\mathcal{F}_Q(v-u)-\frac{v-u}{Q^2+(v-u)^2}\right\}.
\end{align}
The function $\mathcal{F}_Q$ is defined as {\small
\begin{align}
\mathcal{F}_Q(v) = -\frac{i}{4}\left\{
\psi\left(\frac{Q+iv}{4}\right)-\psi\left(\frac{Q-iv}{4}\right)-\psi\left(\frac{Q+2+iv}{4}\right)+\psi\left(\frac{Q+2-iv}{4}\right)
\right\}.
\end{align}
}

\noindent The third and most complicated piece is computed in the same
way as in \cite{Balog:2010vf} by exploiting the relation to the
Heisenberg model discussed above. Here we only quote the
corresponding result
\begin{align}
\delta \mathcal{R}^{(3)}
&=~\frac{1}{\pi}\sum_{Q=1}^{\infty}\int_{-\infty}^{\infty}{\rm
d}v\,  Y^o_{Q+1}(v)\left\{-\mathcal{F}_Q(v-u) + \frac{(v-u)^2-
Q^2}{(v-u) \left((v-u)^2 + Q^2\right)}\right\}.
\end{align}
Finally, summing the corrections up, one can show that the
resulting expression equals to
\begin{align}
\delta\mathcal{R}=
\frac{1}{2\pi}\sum_{Q=1}^{\infty}\int_{-\infty}^{\infty} {\rm
d}v\, \frac{g^{2J}}{(v^2 + Q^2)^J} \frac{\partial}{\partial v}
\left[\frac{T_{Q,1}(v|u)^2}{ S_0(v,u)}\right],
\end{align}
which is in agreement with the proposal by \cite{BJ09}. The
quantities $T_{Q,1}(v|u)$ and $S_0(v,u)$ are given in appendix A.
The expression for $\delta\mathcal{R}$ can be rewritten as
\begin{align}
\delta \mathcal{R} &=
-\frac{1}{2\pi}\sum_{Q=1}^{\infty}\int_{-\infty}^{\infty} {\rm
d}v\,  \left[\frac{d Y^o_Q(v)}{dv} + J
\frac{2v}{Q^2+v^2}Y^o_Q(v)\right]= JP_{\rm LO}\, .
\end{align}
 Thus, at leading order
the corrected Bethe Ansatz equation takes the form
\bea\label{pertmomentum} \frac{2\pi n}{J} =p+P_{\rm LO}(p)\, .
\eea
   This provides a strong argument
for compatibility of the exact Bethe equations with the definition
of the exact, {\it i.e.} non-asymptotic momentum. Equation
(\ref{pertmomentum}) can be solved perturbatively by assuming that
the asymptotic value of $p$ is shifted $p\to p+g^{2J}\delta p$ but
we will not pursue finding an actual solution any further. Rather,
we will continue with model II.

\section{Finite-size corrections to the dispersion: Model II}

In this section we will discuss both the leading and the
next-to-leading order finite-size (wrapping) correction to the
energy in model II.

\subsection{LO wrapping correction to the energy and momentum}
In model II the periodicity conditions for bosons are twisted by
the total momentum of a state; $\alpha=p/2$. Using the
corresponding twisted transfer matrices one finds that for a
single excitation in the limit $g\to 0$ the leading term of
$Y^{o}_Q(v)$ is given by
\begin{align} Y^{o}_Q(v) = \frac{g^{2J+4}}{(v^2 +
Q^2)^{J+2}} \frac{16}{(1+u^2)^2}\frac{
Q^2(Q^2-1+v^2-u^2)^2}{[(Q-1)^2+(v-u)^2][
(Q+1)^2+(v-u)^2]}\, .
\end{align}
The expansion of $Y^{o}_Q$ starts at order $g^{2J+4}$, which also
defines the lowest order of the corresponding wrapping correction
to the Bethe Ansatz energy. Performing a similar computation of
$Y^{o}_Q$ for the multi-particle case, one also finds that in both
level-matched and level-unmatched cases the expansion of $Y^{o}_Q$
starts from the same order $g^{2J+4}$. This behavior should be
contrasted with the one for $Y^{o}_Q$-functions in model I based
on the untwisted transfer matrix. There we found that for the
level-unmatched case the corresponding expansion starts at lower
order $g^{2J}$, irrespective of the number of particles being even
or odd, while for the level-matched case the leading term is of
order $g^{2J+4}$.

\smallskip

In this context it is interesting to recall that in
\cite{Lukowski:2009ce} an analytic continuation of the leading
wrapping correction  from even to odd number of particles was
proposed. This continuation\footnote{Further subtleties concerning
the analytic continuation from even to odd number of particles
have been discussed in \cite{Gomez:2008hx,Gunnesson:2009nn}, where
it was argued that the proposal by \cite{Lukowski:2009ce}
corresponds to one $(-)$ of the two possible prescriptions for
analytic continuation of the harmonic sums with negative values of
indices arising on the gauge theory side.} amounts to modifying
the (untwisted) transfer matrix $T_{Q,1}$ in such a way (by
multiplying certain eigenvalues of the corresponding monodromy
matrix by extra factors of $i$) that in the level-unmatched case
for odd number of particles the expansion of the wrapping
correction starts  starts from $g^{2J+4}$, {\it i.e.} as for
physical states. This modification was proposed to describe an
unphysical magnon with momentum $p=\pi$. It is easy to see that
our construction of the twisted transfer matrix agrees and
correctly generalizes the proposal \cite{Lukowski:2009ce} for the
case of an off-shell theory with any number of excited states and
for any value of the coupling constant\footnote{The twist must
necessarily be $p$-dependent. Indeed, if one admits that in the
single magnon case the constant twist in \cite{Lukowski:2009ce} is
valid not only for $p=\pi$ but rather for all values of $p$, then
for the physical state with $p=0$ one finds a non-vanishing energy
correction which is nonsensical because the corresponding state
being the susy descendant of the BPS vacuum is not renormalized.}.

\smallskip

For a generic rapidity $u$, the resulting wrapping corrections are
given in appendix C. Here we present the wrapping correction on
solutions of the Bethe equations. Similar to model I, we find that
upon substituting the solution of the Bethe equation
$u=\cot\frac{\pi n}{J}$, the contribution of polygamma-functions
in the energy correction completely drops out. This again results
in a compact form of the wrapping correction, given solely in
terms of $\zeta$-functions
\begin{align}\label{eqn;ExactWrappingEonshellII}
E_{\rm LO}^{\rm II}(J) =&~
g^{2J+4}\left[\,\sum_{m=1}^{\left\lfloor
\frac{J}{2}\right\rfloor+1}C_m(J) \zeta (2 J-2 m+1) -
8\sin^4\tfrac{p}{2} \,
\frac{\Gamma(J+\frac{3}{2})}{\Gamma(J+2)}\frac{\zeta(2J+1)}{\sqrt{\pi}}
\right] \, .
\end{align}
Here the coefficients $C_m(J)$ written in terms of particle momentum take the form
\begin{align}
C_m(J)=&(-1)^{m} \frac{16}{\sqrt{\pi }} \frac{\Gamma (J-m+1) \Gamma \left(J-m+\tfrac{3}{2}\right)}{\Gamma (J+2) \Gamma (J-2 m+2)}  \sin ^{2 m+3}\left(\tfrac{p}{2}\right) \times \\
& \hspace{60pt} \times \Big[ m \sin(m+\tfrac{1}{2})p - (J+1) \cos \tfrac{p}{2} \sin mp\Big]. \nonumber
\end{align}
The above formula holds for $J>2$ as for $J=2$ it contains a
divergent term originating from $\zeta(1)$. For this case $E_{\rm
LO}^{\rm II}(2)$ can be computed directly and the results is
\begin{align}
E_{\rm LO}^{\rm II}(2) = 2 \sin ^6\frac{p}{2} (2+\cos p) \zeta(3) - \frac{5}{2} \sin ^4\frac{p}{2}\zeta (5).
\end{align}
It is readily seen that for $p=0$ the wrapping correction vanishes.

\smallskip

Next we present the leading finite size correction to the momentum.
From the identity {\smallskip
\begin{align}
\partial_v Y^{o}_Q(v)+\partial_u Y^{o}_Q(v) + \frac{2v(J+2)}{Q^2+v^2}
Y^{o}_Q(v) +
\left[\frac{4u}{1+u^2}-\frac{4(v-u)}{Q^2-1+v^2-u^2}\right]Y^{o}_Q(v)=0,
\end{align}
}
\renewcommand{\baselinestretch}{1.3}

\noindent it is easily shown that
\begin{align}
\label{PEII} (J+2) P^{\mathrm{II}}_{\rm LO} = -\,\partial_u
E^{\mathrm{II}}_{\rm LO} - \frac{4u}{1+u^2} E^{\mathrm{II}}_{\rm
LO} - 4\mathcal{I}_1,
\end{align}
where
\begin{align}
\mathcal{I}_1(J) = \frac{1}{2\pi}\sum_{Q=1}^{\infty} \int dv ~
\frac{v-u}{Q^2-1+v^2-u^2}Y^0_{Q}(v).
\end{align}
The integral $\mathcal{I}_1$ can be computed explicitly and it is given in appendix
\ref{section;Integrals}.
\smallskip

Using the explicit result for $\mathcal{I}_1$, it is now straightforward to find the expression for the wrapping
corrected momentum by differentiating the wrapping correction to the energy. We define the coefficients $D_m(J)$ to be
{\small \begin{align} D_m(J)=&
\frac{2^{2m-2J+3}(-1)^m\Gamma(2J-2m+2)}{\Gamma(J+3)\Gamma(J-2m+2)}\left(\sin\frac{p}{2}\right)^{2m+4}\nonumber\\
&\left((J-m+1)(2m+1)\sin(n+1)p\,\sin\frac{p}{2} -
2(J+2)m\sin\frac{p}{2} \cos(n+\frac{1}{2})p \right),
\end{align}
}and then $P^{\mathrm{II}}_{\rm LO}$ becomes {\small\begin{align}
P^{\mathrm{II}}_{\rm LO} =& \frac{\left(\sin\frac{p}{2}\right)^{J+2}}{2^{J-1}} \left[ \frac{\sin \frac{Jp}{2} A(J+2)}{4i^{2-J}(J+2)!} + \frac{\sin \frac{(J-1)p}{2}\sin \frac{p}{2} A(J+1)}{2i^{1-J}(J+1)!} + \frac{\sin \frac{(J+2)p}{2}\sin^2\frac{p}{2} A(J)}{i^{-2-J}J!}\right. \nonumber\\
&\left. + 2\frac{\sin \frac{(J+1)p}{2}\sin^3
\frac{p}{2}A(J-1)}{i^{-3-J}(J-1)!}
\right]-\sum_{m=0}^{\lfloor\frac{J+1}{2}\rfloor}D_m(J)\zeta(2J-2m+1).
\end{align}
}Notice that, in contrast to the wrapping correction to the
energy, the polygamma functions here do not cancel on a  solution
of the Bethe equation.

\subsection{Quantization of momentum and exact Bethe equation}
Once again, following \cite{Balog:2010vf,Balog:2010xa}, it is
possible to establish a connection between the leading order of
$Y^{o}_Q$ and the transfer matrix of the Heisenberg spin chain.
For a single excitation with rapidity $u$ we define the following
transfer matrix
\begin{align}
\label{tQ} t_Q(v) = (Q+1)(Q(Q+2)+v^2-u^2)\, .
\end{align}
It is readily checked that it satisfies the equation
\begin{align}
t_{Q}(v+i)t_{Q}(v-i) = t_{Q+1}(v)t_{Q-1}(v) +
t_0(v+(Q+1)i)t_0(v-(Q+1)i)\, .
\end{align}
The associated Y-functions are again defined as
\begin{align}
y_Q(v) = \frac{t_{Q+1}(v)t_{Q-1}(v)}{t_{Q}(v+i)t_{Q}(v-i) -
t_{Q+1}(v)t_{Q-1}(v) } \, .
\end{align}
Explicitly,
$$
y_Q(v)=Q(Q+2)\frac{[Q^2-1+v^2-u^2][(Q+2)^2-1+v^2-u^2]}{[(Q+1)^2+(v-u)^2][(Q+1)^2+(v+u)^2]}\,
.
$$
These functions satisfy the functional relation
\begin{align}
y_Q(v+i)y_Q(v-i) = [1+y_{Q+1}(v)][1+y_{Q-1}(v)].
\end{align}
Obviously, each $y_{Q}(v)+1$ has four roots $\pm u_Q\pm i$, that
is
\begin{align}
y_Q(\pm u_Q\pm i)=-1,
\end{align}
where $u_Q=\sqrt{u^2-Q^2-2Q}$.

In terms of the quantities introduced above, the leading order of
$Y^{o}_Q(v)$ can be conveniently written as
\begin{align}\label{eqn;Y0inXXX}
Y^o_Q(v) = \frac{16
g^{2J+4}}{(v^2+Q^2)^{J+2}}\frac{j_Q}{(1+u^2)^2}\, ,
\end{align}
with
\begin{align}
j_Q =
\frac{t_{Q-1}(v)^2}{\tau(v+(Q+1)i)\tau(v+(Q-1)i)\tau(v-(Q-1)i)\tau(v-(Q+1)i)}\,
,
\end{align}
where \bea \tau(v)=v-u\, . \eea

\smallskip

Expanding $Y_{M|vw}^{o}$-functions at weak coupling, we find $Y_{M|vw}^{o}(v)=y_M(v)$. We would like
point out that curiously enough $y_M$ coincides precisely
with the leading term in the expansion of $Y_{M|vw}^{o}$ for the
case of two particles with the level-matching condition imposed,
{\it cf.} equation (3.11) in \cite{Arutyunov:2009ax}; in
particular, each $y_M$ has four zeroes at weak coupling.
Furthermore, the transfer matrix $t_Q(v)$ given by equation (\ref{tQ})
which we used to express the asymptotic Y-functions for a {\it
single} and, therefore, unphysical magnon, coincides with the
two-particle transfer matrix of the Heisenberg model
$$
t_Q^{2pt}=(Q+1)[Q(Q+2)+(v-u_1)(v-u_2)]
$$
with the level-matching condition $u_2=-u_1$ imposed.

\smallskip

Using this relationship with the Heisenberg model, one can repeat
the computation of the leading correction to the asymptotic Bethe
equations. This time we find \bea \nonumber \delta {\cal
R}=\frac{1}{\pi}\sum_{Q=1}^{\infty}\int {\rm
d}v\Big[\frac{v-u}{(Q-1)^2+(v-u)^2}+\frac{v-u}{(Q+1)^2+(v-u)^2}-\frac{v+u}{Q^2-1+v^2-u^2}\Big]Y_Q^{o}
\, . \eea The relation with to $P_{\rm LO}$ is
\bea\label{eqn;deltaRII} \delta {\cal R}=  (J+2)P^{\mathrm{II}}_{\rm
LO}+2\mathcal{I}_1\, . \eea Thus, for model II the
leading correction to the asymptotic Bethe equations does not
agree with the corresponding correction to the momentum, as
defined by equation (\ref{standardmomentum}). This result suggests that
in a twisted theory the expression for the actual momentum
requires a modification of expression (\ref{standardmomentum})  in
order to agree with the correction to the Bethe ansatz. It would
be interesting to find its non-perturbative definition.


\subsection{Comparing energies of $\alg{sl}(2)$ and $\alg{su}(2)$ states}

It turns out that the momentum twisted versions of the
$\alg{sl}(2)$ and $\alg{su}(2)$ transfer matrix are related in a
very direct manner. In the $\alg{su}(2)$ sector we will denote the
different momentum twists by $+$ and $-$ in each factor, writing
$T_{Q,1}^{\alg{su}(2)\pm}$ for a transfer matrix twisted with
$\alpha = \pm p/2$. As discussed before, for $\alg{sl}(2)$ this
distinction is not needed as the direction of the twist is
inconsequential and we will simply write $T^{\alg{sl}(2)}_{Q,1}$.
Expanding the transfer matrices at weak coupling, one finds at
leading order the following very simple relations
\begin{equation}
T_{Q,1}^{\alg{sl}(2)} = \left(\frac{g^2}{Q^2+u^2}\right)^{\mp
\frac{1}{2}} T_{Q,1}^{\alg{su}(2)\pm}.
\end{equation}
Now we know that a similar prefactor occurs in the expansion of the
$Y_Q$-function (\ref{genericY}),
\begin{equation}
Y_Q = \left(\frac{g^2}{Q^2+u^2}\right)^J T^{\ell}T^{r}
S_{\sl(2)}^{Q1_*} + \ldots \, ,
\end{equation}
which means that we can \emph{effectively} absorb this prefactor in
a shift of  $J$ depending on our choice of twists. In order to
indicate this clearly, we write $Y_Q^{\alg{sl}(2)}$ for the
momentum twisted Y-function in the $\alg{sl}(2)$ sector, and
$Y_Q^{\alg{su}(2)\pm\pm}$ for the Y-function in the $\alg{su}(2)$
sector, where the twists are taken $\pm p/2$ in the left and right
factor, respectively. We then find the relations
\begin{equation}
Y_Q^{\alg{sl}(2)}(J) = Y_Q^{\alg{su}(2)\pm\mp}(J) =
Y_Q^{\alg{su}(2)++}(J+1) = Y_Q^{\alg{su}(2)--}(J-1).
\end{equation}
Indicating the twists in differently twisted theories  by
subscript signs, these relations imply that the LO wrapping
corrections to the energy of the gauge theory operators of the
schematic form $(\gamma^n D Z^J)$, $(\gamma^n_{\pm\mp}\Phi Z^J)$,
and $(\gamma^n_{\pm\pm}\Phi Z^{J\pm1})$ all coincide. Here
$\gamma$ is an element of a representation of the orbifold group,
appropriate for the field-theoretic realization of the twist, and
the integer $n$ labels the $n$-th twisted sector in the orbifold
theory. Finally, the twists $++$ and $--$ also affect the
$\alg{su}(2)$ Bethe equations, {\it cf.} equation (\ref{1magnonsu2}).
Now we immediately see that these twists can be effectively
compensated by shifting $J\to J\pm 1$, leading to the one and the
same Bethe equation $e^{ipJ}=1$. Therefore, the theories obtained
by doing the various twists are directly related by a simple shift
in $J$ in the $\alg{su}(2)$ sector.

\smallskip

Finally we would like to note that there exists a curiously simple
relationship not only between the leading order but also the full
twisted transfer matrices $T_{Q,1}^{\alg{su}(2)\pm}$ and $T_{Q,1}^{\alg{sl}(2)}$. Namely,
\bea T_{Q,1}^{\alg{su}(2)+}T_{Q,1}^{\alg{su}(2)-}={T_{Q,1}^{\alg{sl}(2)}}^2 \, .\eea
An analogous relation holds in the Konishi case, where it was observed that for level matched states $T_{Q,1}^{\alg{sl}(2)}$ and $T_{Q,1}^{\alg{su}(2)}$ coincide \cite{Arutyunov:2009ax}. Of course this agrees with the above relation, since for level matched states the twist is trivial\footnote{For a single magnon the level matched equivalence is completely trivial.}.

\subsection{NLO wrapping correction to the energy}

To describe the next-to-leading order wrapping correction all quantities in (\ref{FSE}) must be carefully expanded to two orders higher in $g$. In what follows we will discuss the expansion of
the different terms separately and give their contribution to the
energy.

\smallskip

First we consider the term in the integral that is not related to
the Y-function
\begin{align}
\frac{d\tilde{p}^Q}{dv} = 1 + 2g^2\frac{v^2-Q^2}{(v^2+Q^2)^2}
+\mathcal{O}(g^4).
\end{align}
This obviously gives rise to the contribution
\begin{align}
E_{\rm NLO}^{(1)} \equiv 2g^2 \sum_{Q=1}^{\infty}\int dv
\frac{v^2-Q^2}{(v^2+Q^2)^2}Y^0_{Q}(v).
\end{align}
This term can be written as the sum of two terms
\begin{align}
E_{\rm NLO}^{(1)}(J) =  \frac{\mathcal{I}_2(J)}{(u^2+1)^2} - 2
E_{\rm LO }^{\mathrm{II}}(J+1)
\end{align}
The new integral $\mathcal{I}_2$ can be expressed in terms of the previously computed wrapping corrections. The explicit expressions in this integral are presented in section \ref{section;Integrals}.

\smallskip

Next we focus on the term $\log (1+Y_Q(v|\hat{u}))$. This term
depends both implicitly (through the Bethe root $\hat{u} = u +
g^2\delta u$) and explicitly on $g$. Expanding this gives
\begin{align}
\log (1+Y_Q(v|\hat{u})) = Y^{\rm LO}_Q(v|u) + Y^{\rm LO,2}_Q(v|u)
+ Y^{\rm NLO}_Q(v|u) + \mathcal{O}(g^{2J+8}),
\end{align}
where $Y^{\rm LO}_Q(v|u)$ is of order $g^{2J+4}$ and $Y^{\rm LO
,2}_Q(v|u), Y^{\rm NLO}_Q(v|u)$ are of order $g^{2J+6}$. Here
$Y^{\rm NLO}_Q(v|u)$ is obtained by expanding the Bethe root
$\hat{u} = u + g^2\delta u$ and is  expressed as
\begin{align}
Y^{\rm NLO}_Q(v|u) &\equiv g^2\,\partial_u Y^{\rm LO}_Q(v|u)\,
\delta u .
\end{align}
The resulting contribution to the NLO wrapping correction from
this piece is consequently given by
\begin{align}
E_{\rm NLO}^{(2)}(J) =g^2 \sin p~ \frac{d}{du}E^{\mathrm{II}}_{\rm
LO }(J),
\end{align}
where we used (\ref{rapidity}) to express $\delta u = \sin p$.

\smallskip

Subsequently, we turn our attention to $Y^{\rm NLO}_Q(v|u)$. It is
useful to split this function into two parts {\small
\begin{align}
Y^{\rm NLO}_{rat}=&2 g^2 \left[ \frac{4
v^2}{\left(Q^2+v^2\right)^2} -\frac{2 v u}{\left(u^2+1\right)
\left(Q^2+v^2\right)} -
\frac{4}{\left(u^2+1\right) \left(Q^2+v^2\right)}+ \frac{J \left(v^2-Q^2\right)}{\left(Q^2+v^2\right)^2}+\right. \nonumber\\
&\quad~~ \left. +\frac{4}{u^2+1} - \frac{4
\left(Q^2-1\right)}{\left(Q^2+v^2\right)
\left(1-Q^2-v^2+u^2\right)} - \frac{1}{Q^2+v^2} -
\frac{4}{\left(u^2+1\right)^2}\right] Y^{0}_Q(v),\\
Y^{\rm NLO}_{\psi}=& g^2
\left(\frac{H_{-\frac{Q+iv}{2}}+H_{\frac{i v-Q}{2}}+H_{\frac{Q-i
v}{2}}+H_{\frac{Q+i v}{2}}}{u^2+1}\right)Y^{0}_Q(v).
\end{align}
}The contribution $E_{\rm NLO}^{(3)}$ of the first term $Y^{\rm
NLO}_{rat}$ can be related, by some partial integrations to
integrals that were encountered before. In this respect it is
computed in a straightforward manner. The relevant integrals are
collected in appendix D. The second term comes from expanding the
dressing phase, as presented in appendix A. Note that we here used
harmonic numbers $H_n$ rather than the digamma function $\psi(n)$.
They are related via $H_n= \gamma + \psi^{(0)}(n+1)$. Appendix D
is devoted to computing the corresponding contribution which
we denote as $E_{\rm NLO}^{(4)}$.

\smallskip

Finally, we consider the term involving the dispersion relation
$\mathcal{E}(p)$ from equation (\ref{FSE}). Since the momentum also
receives a wrapping correction, {\it i.e.}
\begin{align}
p \rightarrow p + g^{2J+4}\delta p,
\end{align}
the asymptotic energy ${\cal E}(p)$ also gets corrected
\begin{align}
\mathcal{E}(p) &= \sqrt{1+4 g^2\sin^2\frac{p + \delta p}{2}} =
\sqrt{1+4 g^2\sin^2\frac{p}{2}} + g^{2J+6}\sin p\, \delta p  +
\mathcal{O}(g^{2J+8}).
\end{align}
The TBA approach relates $\delta p$ to the correction to the
asymptotic Bethe equation $\delta\mathcal{R}$ as
\begin{align}
\delta p = -\frac{\delta \mathcal{R}}{L_{\scriptsize TBA}}.
\end{align}
Now it is time to recall that for model II the correction $\delta
p$ is not proportional to $P^{\mathrm{II}}_{\rm LO}$. On top of this,
from our TBA considerations in section 4 we found that $L_{TBA} =
J+1$, which results in the following expression for $\delta p$,
{\it cf.} formula (\ref{eqn;deltaRII}),
\begin{align}
\label{nontrivialp} \delta p = -\frac{J+2}{J+1} P_{\rm
LO}^{\mathrm{II}} - \frac{2}{J+1}\mathcal{I}_1(J).
\end{align}
Before continuing with our discussion of the next-to-leading order
correction to the energy, let us mention that there is a
consistency check on our result. Each of the above discussed
intermediate terms contain polygamma functions. However, just as
for the first wrapping correction, we can expect them to cancel in
the final result. It turns out that the contribution for polygamma
functions indeed exactly cancel out for our rather non-trivial
result for $\delta p$.

\subsection*{Collected results}

We have explicitly computed the next-to-leading order wrapping
corrections for states with $J=2,3,4$ and for one state with
$J=6$. We found that in each case the result is a combination of
$\zeta$-functions and multiple (double) zeta-functions (see
appendix D). We will present the results with all double zeta
values expressed in terms of usual $\zeta$-functions. Below we
summarize our results.

\subsection*{$\mathbf{J=2}$}

For $J=2$, the Bethe equations admit only the solutions $p=0,
\pi$, i.e. $u=\infty$ and $u=0$. The contribution for $p=0$ vanishes, and the NLO wrapping
correction for $p=\pi$ is found to be
\begin{align}
\frac{E_{\rm NLO}^{\rm II}(2)}{g^{10}}=-4 \zeta (3)-\frac{3}{2}\zeta
(3)^2-5 \zeta (5)+\frac{105}{8}\zeta (7).
\end{align}

\smallskip

\subsection*{$\mathbf{J=3}$}
For $J=3$ the Bethe roots are
$u=\pm\frac{1}{\sqrt{3}}$. Both rapidities yield the same wrapping
correction
\begin{align}
\frac{E_{\rm NLO}^{\rm II}(3)}{g^{12}}=\frac{243}{512}\zeta
(3)^2 - \frac{243}{128}\zeta (5) - \frac{405}{512}\zeta (3) \zeta
(5)-\frac{21735}{4096} \zeta (7) + \frac{567}{64}\zeta (9).
\end{align}

\smallskip

\subsection*{$\mathbf{J=4}$}
For $J=4$ there are three Bethe roots $u=0,u=\pm1$. For $u=0$ we
find
\begin{align}
\frac{E_{\rm NLO}^{\rm II}(4)}{g^{14}}=&\zeta (3)^2 -\frac{5}{32}\zeta
(5)^2 + \zeta (3) \left(\frac{1}{4}\zeta (5)-\frac{7}{4}\zeta
(7)\right)\\ \nonumber &\hspace{2.5cm} -\frac{9}{2}\zeta
(5)-7 \zeta (7)-\frac{21}{4}\zeta(9) + \frac{1155}{64}\zeta
(11)\, ,
\end{align}
while for $u=\pm 1$ the result is
\begin{align}
\frac{E_{\rm NLO}^{\rm II}(4)}{g^{14}}=&-\frac{1}{32}\zeta
(3)^2 - \frac{5}{256}\zeta (5)^2 +\zeta (3) \left(\frac{7}{32}\zeta (5)-\frac{7}{32}\zeta (7)\right)\\ \nonumber
&\hspace{2.5cm} +\frac{9}{64}\zeta (5) - \frac{119}{640}\zeta(7)-\frac{2037}{640}\zeta(9) + \frac{1155}{256}\zeta(11)\, .
\end{align}

\smallskip

\subsection*{$\mathbf{J=6}$}
Finally we computed the NLO wrapping correction for $J=6,u=0$ and
found
\begin{align}
\frac{E_{\rm NLO}^{\rm II}(6)}{g^{16}}=& -\frac{5}{2}\zeta (7)-\frac{7}{512}\zeta (7)^2 - \zeta (5) \left(\frac{3}{32}\zeta(7) +
\frac{15}{64}\zeta (9)\right)-\frac{51}{8}\zeta (9)-\frac{957}{128}\zeta (11)+\\
&+\zeta (3)\left(\frac{3}{4}\zeta (5)+\frac{9}{16}\zeta(7) + \frac{3}{8}\zeta(9)-\frac{99}{64}\zeta(11)\right) -
\frac{1287}{256}\zeta(13)+\frac{45045}{2048}\zeta(15).\nonumber
\end{align}
One can see from these explicit results that the term of maximum
transcendentality is always of degree $2J+3$. In case of the
leading order wrapping correction this was $2J+1$.

\smallskip

In the above, we have chosen to express the double zeta-functions
in terms of usual $\zeta$'s and products thereof, but we can
also attempt to go the other way around and express the NLO
correction purely in terms of double zeta-functions; this turns
out to be possible. As a matter of fact, all the wrapping
corrections are of the following form
\begin{align}
E_{\rm NLO}^{\rm II} = g^{2J+6}\sum_{i>j}A_{ij}~\zeta(i,j).
\end{align}
For instance, for $J=2$ this formula explicitly reads
\begin{align}
E_{\rm NLO}^{\rm II}(2) =& -4\zeta(2,1)
-10(3\zeta(4,1)+\zeta(3,2))-\frac{3}{2}(16\zeta(5,1)+9\zeta(4,2))+\nonumber\\
&+\frac{105}{8}(20\zeta(6,1)+10\zeta(5,2)+3\zeta(4,3)).
\end{align}
Since the leading order wrapping correction was of the form
\begin{align}
E_{\rm LO} = g^{2J+4}\sum_{i}A_{i}~\zeta(i),
\end{align}
the general structure that is expected to be found for higher order wrapping corrections will be ordered sums of multiple
zeta values.

\smallskip

We close this section with the following curious observation. As
we have seen, the relation between $\delta p$ and $P_{\rm LO}$ is
given by equation (\ref{nontrivialp}). It is interesting to ask what
we would find for the NLO energy correction if we would assume
that $\delta p$ and $P_{\rm LO}$ would be related in the standard
fashion: $\delta p=-P_{\rm LO}$. It turns out that the
contribution of $\psi$-functions also cancels out completely  in
the expression for $E_{\rm NLO}$! The final result differs,
however, from $E_{\rm NLO}^{\rm II}$, but only in terms linear in
$\zeta$-functions. For example, for $J=4,u=1$, the difference is
$\frac{21 \zeta (7)}{640}-\frac{21 \zeta (9)}{320}$. In fact, we
have found that the following linear combination
\begin{align}
P_{\rm LO}^{\mathrm{II}}+2\mathcal{I}_1(J)
\end{align}
is ``$\psi$-free" and, therefore, being added with arbitrary
coefficient $\alpha$ to our original expression for $\delta p$
does not spoil the ``$\psi$-free" property: \bea \delta
p=-P_{\rm LO}^{\rm II} + \frac{\alpha}{J+1}(P_{\rm LO}^{\mathrm{II}} + 2\mathcal{I}_1(J))\, .\eea When
$u=0$ both choices for $\delta p$ give rise to the same NLO
correction to the energy. The implications of these observations
are not clear to us.

\section{Finite-size corrections in $\beta$-deformed and physical orbifold theories}

In this section we apply the generalized L\"uscher formulae to
compute the leading finite-size corrections to the energies of
one- and two-particle excited states from the $\su(2)$ and
$\sl(2)$ sectors in $\beta$-deformed theories, and to two-particle
states in the $\sl(2)$ sector of a generic ${\mathbb
Z}_S$-orbifold. The $Y_Q(v)$functions needed to compute $E_{\rm
LO}$ according to equation (\ref{eqn;Luscher}) are constructed by
using the twisted transfer matrices and the general formula
(\ref{genericY}). The rest of our considerations follows the same
logic as for models I and II. We also make a comparison between
our results for $\beta$-deformed and orbifold theories for a
single magnon.

\subsection{One-particle $\beta$-deformed states}
For single particle states, we find the following energy corrections for the $\alg{su}(2)$ and $\alg{sl}(2)$ sectors.
\subsubsection*{ $\alg{su}(2)$ sector}
According to our discussion in section 2, a single magnon in the
$\su(2)$ sector of  $\beta$-deformed theory has the following
Bethe equation \bea\label{pbetadeformed} 1=e^{i(p-2\pi \beta)J}
~~~~~~\Longrightarrow~~~~~~p=2\pi \beta+\frac{2\pi n}{J}\, ,~~~~~
n=0,\ldots \Big[\frac{J}{2}\Big]\, .  \eea  For the level-matched
case $n=0$ so that a physical magnon has $p=2\pi \beta$.
 Its
asymptotic energy is
\begin{align}
E = J+\sqrt{1+4g^2 \sin^2 \pi\beta}.
\end{align}
On the gauge theory side an operator corresponding to the physical
magnon is ${\tr}(\Phi Z^J)$.

\smallskip

Let us denote the transfer matrix obtained by twisting the bosonic
eigenvalues (in the auxiliary space) with $\alpha$ as
$T(v|u;\alpha)$. Implementing the twist describing
$\beta$-deformed theory results in the following Y-functions
\begin{align}
Y_Q(v) &=
\frac{g^{2J}}{(Q^2+v^2)^J}\frac{T^{\alg{su}(2)\ell}_{Q,1}(v|u;\beta)T^{\alg{su}(2)\, r}_{Q,1}(v|u;(2J-1)\beta)}{S_0(u,v)}
\, ,
\end{align}
where the coefficient $S_0(u,v)$ is given in appendix A. The
leading term in the weak-coupling expansion for the twisted
transfer matrix in the $\alg{su}(2)$ sector is explicitly given by
{\small
\begin{align}
T_{Q,1}^{\alg{su}(2)}(v|u;\alpha)=&\frac{g}{\sqrt{Q^2+v^2}} \left[
\frac{(Q-1)\left((Q+1)^2+(v-u)^2\right)}{(i (Q-1)+v-u)(i-u)} +
\frac{Q e^{i \pi  \alpha }\sqrt{\frac{u+i}{u-i}} (Q^2+(u-v+i)^2)}{(i (1-Q)-v+u)(i-u)} + \right.\nonumber\\
&\left.\qquad\qquad\quad+ \frac{Q e^{-i \pi  \alpha } (i
(Q+1)-v+u)}{(i-u)\sqrt{\frac{u+i}{u-i}}} + \frac{(Q+1) (i
(1-Q)+v-u)}{i-u}\right].
\end{align}
}This transfer matrix starts at order $g$ which results in the
fact that our Y-function, and corresponding corrections to the
energy and momentum, will be of order $g^{2J+2}$.

\smallskip

It is now straightforward to compute the first wrapping correction
to the energy. We find it to be a sum of $\zeta$-functions
\begin{align}
\label{Esu1magnon} E^{\beta}_{\rm LO}(J) =g^{2J+2}
\sum_{n=1}^{\lfloor\frac{J+1}{2}\rfloor} 16 (-1)^n\frac{\Gamma
(J-n+1) \Gamma \left(J-n+\frac{3}{2}\right)}{\sqrt{\pi}\, \Gamma
(J+1) \Gamma (J-2 n+3)}B^{\beta}_n(J) \zeta (2 J-2 n+1) ,
\end{align}
where the coefficients are given by the following expressions
{\smallskip
\begin{align}
& B^{\beta}_{1}(J) = -\frac{J}{2}\sin(J\pi\beta)\sin^2(\pi\beta)\sin((J-2)\pi\beta),\\
& B^{\beta}_{n>1}(J) = \sin(J\pi\beta) \sin^{2n}(\pi\beta) \Big[
(n-1) \sin ((J-2n)\pi\beta)-J \cos(\pi\beta)\sin((J-2
n+1)\pi\beta)\Big]\nonumber.
\end{align}
}The leading finite-size correction to the energy found above
perfectly agrees with the field-theoretic result obtained in
\cite{Fiamberti:2008sm,Fiamberti:2008sn}. As discussed in the introduction, this is also in agreement with the results of \cite{Gromov:2010dy}.

\smallskip

Special attention has to be paid to the particular case  of
rational $\beta$, $\beta = \frac{n}{J}$ with $n=0,1,\ldots, J-1$.
For these values of the deformation parameter, the naive lowest order in
the weak-coupling expansion of the transfer matrix
$T_{Q,1}^{\alg{su}(2)}(v|u;(2J-1)\beta)$ vanishes and the expansion really starts two powers of $g$ higher. Bearing in mind that the rapidity $u$ is $g$-dependent via (\ref{rapidity}) with
momentum $p=2\pi\beta$, it is readily checked that the functions
$Y_Q$ for these special values of $\beta$ are absolutely the same
as for generic $\beta$, up to a shift $J\rightarrow J+1$, so that we find
\begin{align}
E^{\beta=n/J}_{\rm LO}(J) = \left.E^{\beta}_{\rm
LO}(J+1)\right|_{\beta=\frac{n}{J}}.
\end{align}
In particular this implies that for these special values of $\beta$  the expansion of the
$Y_Q$-functions starts at order $g^{2J+4}$.

\subsubsection*{$\alg{sl}(2)$ sector}
In the $\alg{sl}(2)$ sector the single magnon Bethe equations are
$e^{ipJ} =1$ so that
\begin{align}
p &= \frac{2\pi n}{J}\, , ~~~~~~~~ n=0,1,\ldots J-1.
\end{align}
The physical state must be level-matched leaving only the solution
$p=0$. In the $\sl(2)$ sector of  $\beta$-deformed theory the left
and right transfer matrices $T_{Q,1}$ are twisted with $\alpha = 0$ and $\alpha =
2J\beta$, respectively. Computing then the
corresponding twisted $Y_Q$-functions, we find that they all
vanish for $p=0$, resulting in a vanishing wrapping correcting to
the energy.

\subsection{Two-particle $\beta$-deformed states}

The twist should correctly account for $\beta$-deformation of
$\mathcal{N}=4$ SYM for any number of excitations; in this section
we will briefly investigate some two-particle states. In
particular, we reproduce the wrapping energy correction found in
\cite{Fiamberti:2008sm,Ahn:2010yv} for the Konishi state in the
$\alg{su}(2)$ sector of $\beta$-deformed theory, and investigate
two-particle states in the $\sl(2)$ sector for which no
field-theoretic results are known so far.

\subsubsection*{$\alg{su}(2)$ sector}
The perturbative computation of the anomalous
dimension of $\beta$-deformed $\alg{su}(2)$ Konishi-like states
\cite{Fiamberti:2008sm} agrees perfectly with results coming from
the asymptotic Bethe equations up to three loops. At four loops, there is a discrepancy
arising due to wrapping effects, which should accounted for via energy corrections computed from the L\"{u}scher
formulae. In \cite{Ahn:2010yv} a modified S-matrix was used to
obtain an expression for the wrapping correction to the energy,
which is in agreement with the explicit perturbative results.

\smallskip

The modification of the S-matrix proposed in \cite{Ahn:2010yv}
amounts to giving the matrix elements of the S-matrix for
fermionic states a $\beta$-dependent phase:
\begin{equation}
\label{eq:modSnepo} \mathcal{S}^{(l1)^{j1}}_{\hspace{14pt}_{j1}}
\rightarrow \begin{cases}
e^{i \pi \beta} \mathcal{S}^{(l1)^{j1}}_{\hspace{14pt}_{j1}},& j = 2l+1,\ldots,3l,  \\
e^{-i \pi \beta}  \mathcal{S}^{(l1)^{j1}}_{\hspace{14pt}_{j1}}, & j = 3l+1,\ldots,4l.  \\
\end{cases}
\end{equation}

\noindent This is obviously different from the twisting procedure
we have proposed above, and therefore we cannot a priori expect
these two different approaches to agree. However, the
$\alg{su}(2)$ Konishi state  is special, because it has
particularly symmetric quantum numbers; $J=M$. This means that for
such a state our twist amounts to a left twist by $M$ and a right
twist by $2J-M=M$. For such a symmetric twist it is not hard to
see that it can effectively be absorbed in the \emph{unmodified}
S-matrices, exactly in the fashion indicated above. This means, in
particular, that for the Konishi state the leading wrapping
correction computed from the twisted Y-functions must naturally
agree with the result of \cite{Ahn:2010yv}:
\begin{equation*}
E^{\beta}_{\mathrm{Konishi}} = g^8 \left[-54(1+\Delta)^3(-5+3\Delta) \zeta(3) -
360 (1+\Delta)^2 \zeta(5) +
\frac{81(1-3\Delta)^2(1+\Delta)^4}{(1+3\Delta)^2}\right],
\end{equation*}
where $\Delta = \tfrac{\sqrt{5+4\cos{4\pi \beta}}}{3}$. Of course,
we have also verified this agreement by explicit calculation based
on the twisted transfer matrix, both analytically for certain nice
values of $\beta$, and numerically for various random values of
$\beta$.

\smallskip

We would like to emphasize that \emph{only} in the special case of
$J=M$ it is possible to effectively absorb the twist in the
modification of the S-matrix in a natural way, and, therefore,
these different approaches will \emph{only} agree in this special
case.

\subsubsection*{$\alg{sl}(2)$ sector}
Now we apply our approach to analyze the wrapping corrections  in
the $\alg{sl}(2)$ sector of $\beta$-deformed theory. We restrict
ourselves to two-particle states with angular momentum $J$,
corresponding to the gauge-theory operators schematically of the
form $\mbox{tr}(D^2 Z^J)$.

\smallskip

 To obtain the corresponding $Y_Q$-functions, {\it cf.} our earlier discussion in
section 2, we need to twist the right transfer matrix with a
factor of $2J \pi \beta$ and to keep the left transfer matrix
untwisted, \emph{regardless} of the number of excitations.
However, since the Bethe equations in the $\alg{sl}(2)$ sector are
not modified under $\beta$-deformation (\ref{BYsl2}), the standard
level matching condition remains $p=0$. With this twist and the
level matching condition $u_1=-u_2=u$ taken into account, the left
transfer matrix for two particles take the following form in the
weak coupling limit:
\[
T^{\alg{sl}(2) \ell}_{Q,1} = \frac{8 g^2 Q
\left(Q^2-1+v^2-u^2\right)}{\left(1+u^2\right)
\left(Q^2+v^2\right) \left((Q-1)^2-2 i (Q-1) v-v^2+u^2\right)},
\]
which for $J=2$ is of course nothing else but the transfer matrix
for the usual Konishi state. On the other hand, due to the twist,
the right transfer matrix starts with a non-trivial zeroth order
term
\begin{equation}
\label{eq:sl22partbetatwist}
T^{\alg{sl}(2)\, r}_{Q,1} = \frac{4 Q \sin^2(\pi  J
\beta)}{3}\frac{\left(1-Q^2 + 3v^2 - 3u^2\right)}{(v+i
(Q-1))^2-u^2},
\end{equation}
This immediately implies that the wrapping correction will show up
at an order $g^2$ lower, except for the special
values\footnote{For these values the twist effectively disappears,
i.e. we return to the undeformed theory.} $\beta = m/J$, $m =
{0,1,\ldots,J}$. Putting these results together in
equation (\ref{genericY}), yields the $Y_Q$-function
\[
Y_Q = \sin^2{(J\pi\beta)}
\frac{g^{2J+2}}{(Q^2+v^2)^{2J+2}}\frac{32 Q^2 \left(u^2+1\right)
\left(Q^2+v^2-u^2-1\right) \left(Q^2-3 v^2+3 u^2-1\right)}{3 f^+_+
f^+_- f^-_+ f^-_-},
\]
where
\[
f^\pm_\mp=(Q \pm 1)^2+(v \mp u)^2.
\]
Thus, we are now able to compute the wrapping corrections to the
energy for two-particle $\alg{sl}(2)$ states with an arbitrary
twist. Evaluating the energy integral for arbitrary $u$, gives a
string of polygamma functions in addition to the expected
$\zeta$-functions. However, substituting the actual solution of
the  Bethe equations, these polygamma contributions cancel out in
a highly non-trivial manner.

\smallskip

Here we explicitly present the wrapping corrections to the energy
for a few values of $J$ where the rapidities take particularly
simple values; since in the small coupling limit $u_{J,n} =
\cot{\frac{\pi n}{J+1}}$, where $n=1,\ldots,\lfloor
\tfrac{J+1}{2}\rfloor$, we restrict ourselves to $J=2,3$ and $5$. The results for the corresponding wrapping
corrections are summarized in Table \ref{tab:sl2res}.

\begin{table}
\begin{center}
 \renewcommand{\arraystretch}{1.5}
\renewcommand{\tabcolsep}{0.2cm}
\begin{tabular}{|c|c|c||c|}
\hline
$J$ & $n$ & $u$ & $E^{\beta}(J)/(g^{2J+2}\sin^2(J\pi\beta))$\\
\hline \hline
2 & 1 & $1/\sqrt{3}$& $  \frac{3}{8}$\\
3 & 1 & $1$& $-\tfrac{1}{12} - \tfrac{1}{3}\zeta(3) + \tfrac{5}{6}\zeta(5)$\\
3 & 2 & $0$& $-6+\tfrac{20}{3}\zeta(3)-\tfrac{5}{3}\zeta(5)$\\
5 & 1 & $\sqrt{3}$& $-\frac{1}{1536}-\frac{1}{192}\zeta (3)+\frac{17}{192}\zeta (5)-\frac{161}{384}\zeta (7)+\frac{21}{32}\zeta (9)$\\
5 & 2 & $1/\sqrt{3}$& $-\frac{81}{512}-\frac{27}{64}\zeta (3)+\frac{27}{64}\zeta (5)+\frac{63}{128}\zeta (7)$\\
5 & 3 & $0$& $-\frac{26}{3}-\frac{4}{3} \zeta (3)+\frac{41}{3} \zeta (5)-\frac{7}{3} \zeta (7)-\frac{21}{16} \zeta (9)$\\
\hline
\end{tabular}
\caption{The leading finite-size correction to the energy for
certain two-particle states from the $\sl(2)$ sector in
$\beta$-deformed theory with $\beta\neq \frac{m}{J}$, $m\in
\mathbb{Z}$. For $J=2$ the leading correction arises at three
loops and for $J=3$ at four. Note that compared to the other two $J=5$ states, the term of maximum transcendentality for $J=5$,$u=1/\sqrt{3}$ is missing.} \label{tab:sl2res}
\end{center}

\end{table}

\subsection{Physical two-particle $\alg{sl}(2)$ orbifold states}
While the Bethe equations for a two-particle state in the
$\alg{sl}(2)$ sector are not modified by the particular
orbifolding corresponding to the twist (\ref{orbi}), the wrapping
corrections to the energy are. To consider
two-particle states on a $\mathbb{Z}_S$-orbifold, we should give
both $\alg{sl}(2)$ transfer matrices the same twist; one of the
$S$-th roots of unity. The explicit expression for the lowest
order term of the twisted transfer matrix has already been given
in equation (\ref{eq:sl22partbetatwist}), where now all we need to do is
replace $2 J \pi \beta$ by $2 \pi m/S$, where $m=0,\ldots,S-1$,
and $m=0$ of course gives the usual $\alg{sl}(2)$ descendant of
the Konishi state. Using formula (\ref{eq:sl22partbetatwist}), the
leading finite-size correction to the energy of any desired state
can be readily computed, here we present the explicit results for
some simple states in Table \ref{tab:sl2orbres}.

\begin{table}
\begin{center}
 \renewcommand{\arraystretch}{1.5}
\renewcommand{\tabcolsep}{0.2cm}
\begin{tabular}{|c|c|c||c|}
\hline
$J$ & $S$ & $u$ & $E^{orb}(J)/(g^{2J}\sin^4( \pi m/S)$\\
\hline \hline
2 & 1 & $1/\sqrt{3}$& $  -\frac{1}{3}$\\
3 & 1 & $1$& $-\frac{1}{18}-\frac{1}{3}\zeta (3)$\\
3 & 2 & $0$& $-1+\frac{16}{9}\log(2)-\frac{1}{3}\zeta (3)$\\
5 & 1 & $\sqrt{3}$& $-\frac{1}{576}-\frac{1}{72}\zeta (3)+\frac{5}{18}\zeta (5)-\frac{35}{36}\zeta(7)$\\
5 & 2 & $1/\sqrt{3}$& $-\frac{3}{64}-\frac{1}{8}\zeta (3)$\\
5 & 3 & $0$& $-\frac{13}{9}+\frac{16}{9}\zeta (3)-\frac{5}{9}\zeta (5)-\frac{35}{144}\zeta(7)$\\
\hline
\end{tabular}
\caption{The leading finite-size correction to the energy for
certain two-particle states from the $\sl(2)$ sector for a
$\mathbb{Z}_S$-orbifold. The leading order correction enters at
order $g^{2J}$. It is interesting to note that compared to the
other two $J=5$ states, the term of maximum transcendentality for
$J=5$,$u=1/\sqrt{3}$ is missing.} \label{tab:sl2orbres}
\end{center}

\end{table}

\subsection{Comparing $\beta$-deformed and off-shell orbifold magnons}
According to equation (\ref{pbetadeformed}), a single $\su(2)$ magnon in
$\beta$-deformed theories has momentum $p=2\pi \beta+\frac{2\pi
n}{J}$. Clearly the case of $\beta$ rational,
$\beta=\frac{m}{J}$, $m\in {\mathbb Z}$, is special, because now
$p=\frac{2\pi(m+n)}{J}$, {\it i.e.} the momentum is essentially
the same as of an {\it off-shell} $\sl(2)$ magnon in orbifold
theory. This suggests the existence of a certain relation between
the $\su(2)$ magnons of $\beta$-deformed theories with
$\beta=\frac{m}{J}$ and $\sl(2)$ magnons of orbifold theory;
their energies must be equal. We could also come to a similar
conclusion by inspecting table \ref{tab:twists}. Indeed, for
$\beta=\frac{m}{J}$ the right twist in the $\su(2)$ sector is
$(2J-M)\pi \beta=2\pi m-\pi \beta M \cong -\pi \beta M$. It is
clear now that if for model II the total momentum is chosen to be
$p=2\pi\beta M$, then the twist in the $\su(2)$ sector of
$\beta$-deformed theory just coincides with the one in model II.

\smallskip

In fact, the equality between the energy of an $\su(2)$ magnon in
 $\beta$-deformed theory with a rational $\beta$ and the energy
of the $\alg{sl}(2)$ magnon in the off-shell theory (with
$\beta=0$) has been already pointed out in
\cite{Gunnesson:2009nn}. There it was argued to arise due to the
symmetry of the spectrum under the shift $\beta\to
\beta+\frac{1}{J}$. In particular, a physical magnon in
$\beta$-deformed theory with $\beta=1/2$ was shown to have the
same leading correction to the energy as a physical magnon with
$p=\pi$ in the undeformed theory
 \cite{Gomez:2008hx,Gunnesson:2009nn,Beccaria:2009hg}. The formulae for the energy
 corrections we have obtained allow us to give a proof of the
 corresponding
 statement for arbitrary $J$.
 Indeed, comparing formula (\ref{Esu1magnon}) taking for $\beta = \frac{n}{J}$
 against the wrapping correction (\ref{eqn;ExactWrappingEonshellII}) in model II for $p=\frac{2\pi n}{J}$ one finds
\begin{align}
E^{\beta}(J+1) = E^{\mathrm{II}}(J).
\end{align}

Actually, the observation made for model II extends\footnote{One
can try to compare $\beta$-deformed theory against model I. Here
one does not expect any obvious relation, but we still find a
certain structural similarity of the leading order corrections to
the energy up to normalization and the piece of maximum
transcendentality
\begin{align}
E^{\beta}(J-1) =  -64 \cos ^4\left(\frac{m \pi }{2 J}\right)
E^{\mathrm{I}}(J) -8 g^{2J} \sin ^4\left(\frac{\pi  m}{J}\right)
\frac{\Gamma \left(J-\frac{1}{2}\right) }{\sqrt{\pi } \Gamma
(J)}\,  \zeta(2J-3).
\end{align}
} to the level of Y-functions. Namely, we have that {\small
\begin{align}\label{compYbetaII}
Y^{\beta}_Q(J+1)& = \frac{\sin ^2(\pi  (J+1) \beta )}{ \sin^2(\pi
\beta )}Y_Q^{\mathrm{II}}(J) \\ \nonumber &\hspace{2cm} +
\frac{\cos(\pi\beta)-\cos(\pi (2
J+1)\beta)}{\sin^3(\pi\beta)}\frac{v-u}{Q^2-1+v^2-u^2}Y_Q^{\mathrm{II}}(J)\,
.
\end{align}
}We see that for $\beta=\frac{n}{J}$ only the first term on
the right hand side of the last expression survives. This also
automatically implies that $P_{\rm LO}^{\beta}$ for
$\beta=\frac{n}{J}$ exactly coincides with $P_{\rm LO
}^{\mathrm{II}}$. Notice that the second term in
equation (\ref{compYbetaII}) coincides with the difference between
$\delta\mathcal{R}$ and $P_{\rm LO}^{\mathrm{II}}$.

\smallskip

Finally, we can speculate on the relevance of the NLO result for model
II in the context of $\beta$-deformed theories. If for model II
the equivalence between the $\sl(2)$ and the $\su(2)$ sectors
continues to hold beyond the leading order, it is probable
that the results for $E_{\rm NLO}^{\rm II}$ we found in section 6
also represent the NLO energy corrections for $\su(2)$ states in
$\beta$-deformed theory with $\beta=\frac{n}{J}$. Coincidence of
$P_{\rm LO}^{\beta}$ with  $P_{\rm LO}^{\mathrm{II}}$ can be
considered as serious evidence in favor of this. However, since
currently we do not know what $\delta {\cal R }$ is in
$\beta$-deformed theory and how it is related to $P_{\rm
LO}^{\beta}$, we can not directly verify the equality of the NLO
energy corrections.

\smallskip

We could also speculate that in $\beta$-deformed theory the situation
might be slightly better, in the sense that possibly $\delta
p=P_{\rm LO}^{\beta}$, in line with our curious observation at the end of section 6. In that case the corresponding energy correction could be easily obtained from our results for $E_{\rm
NLO}^{\rm II}$ and it would differ from the later by terms linear in $\zeta$-functions. However, for the moment this is just a bold speculation.

\section*{Acknowledgements}
We are grateful to Sergey Frolov for numerous important
discussions, and to Zoltan Bajnok, Christoph Sieg and Matteo Beccaria for valuable
comments on the manuscript. G.A. acknowledges support by the
Netherlands Organization for Scientific Research (NWO) under the
VICI grant 680-47-602. The work by M.L. and S.T. is a part of the
ERC Advanced Grant research programme No. 246974, {\it
``Supersymmetry: a window to non-perturbative physics"}.

\bibliographystyle{JHEP}

\newpage

\section{Appendices}

\subsection{Appendix A: Twisted transfer matrix}
The eigenvalue of the twisted transfer matrix for an
anti-symmetric bound state representation with the bound state
number $Q$ is given by the following formula, generalizing the
result of \cite{Arutyunov:2009iq}
\begin{eqnarray}\label{eqn;FullEignvalue}
&&T_{Q,1}(v\,|\,\vec{u})=\prod_{i=1}^{K^{\rm{II}}}{\textstyle{\frac{y_i-x^-}
{y_i-x^+}\sqrt{\frac{x^+}{x^-}}
\, +}}\\
&&
{\textstyle{+}}\prod_{i=1}^{K^{\rm{II}}}{\textstyle{\frac{y_i-x^-}{y_i-x
^+}\sqrt{\frac{x^+}{x^-}}\left[
\frac{x^++\frac{1}{x^+}-y_i-\frac{1}{y_i}}{x^++\frac{1}{x^+}-y_i-\frac{1
}{y_i}-\frac{2i Q}{g}}\right]}}\prod_{i=1}^{K^{\rm{I}}}
{\textstyle{\left[\frac{(x^--x^-_i)(1-x^-
x^+_i)}{(x^+-x^-_i)(1-x^+
x^+_i)}\frac{x^+}{x^-}  \right]}}\nonumber\\
&&{\textstyle{+}}
\sum_{k=1}^{Q-1}\prod_{i=1}^{K^{\rm{II}}}{\textstyle{\frac{y_i-x^-}{y_i-
x^+}\sqrt{\frac{x^+}{x^-}}
\left[\frac{x^++\frac{1}{x^+}-y_i-\frac{1}{y_i}}{x^++\frac{1}{x^+}-y_i-\frac{1}{y_i}-\frac{2ik}{g}}\right]}}
\left\{\prod_{i=1}^{K^{\rm{I}}}{\textstyle{\lambda_+(v,u_i,k)+}}\right.\left.\prod_{i=1}^{K^{\rm{I}}}{\textstyle{\lambda_-(v,u_i,k)}}\right\}\nonumber\\
&&\quad -\sum_{k=0}^{Q-1}\prod_{i=1}^{K^{\rm{II}}}
{\textstyle{\frac{y_i-x^-}{y_i-x^+}\sqrt{\frac{x^+}{x^-}}\left[\frac{x^+
-\frac{1}{x^+}-y_i-\frac{1}{y_i}}
{x^+-\frac{1}{x^+}-y_i-\frac{1}{y_i}-\frac{2ik}{g}}\right]}}\prod_{i=1}^
{K^{\rm{I}}}{\textstyle{\frac{x^+-x^+_i}{x^+-x^-_i}\sqrt{\frac{x^-_i}{x^
+_i}} \left[1-\frac{\frac{2ik}{g}}{v-u_i+\frac{i}{g}(Q-1)
}\right]}}\times\nonumber\\
&&\quad\times
\left\{e^{i\alpha}\prod_{i=1}^{K^{\rm{III}}}{\textstyle{\frac{w_i-x^+-\frac{1}{x^+}
+\frac{i(2k-1)}{g}}{w_i-x^+-\frac{1}{x^+}+\frac{i(2k+1)}{g}}+ }}
e^{-i\alpha}\prod_{i=1}^{K^{\rm{II}}}{\textstyle{\frac{y_i+\frac{1}{y_i}-x^+-\frac
{1}{x^+}+\frac{2ik}{g}}{y_i+\frac{1}{y_i}-x^+-\frac{1}{x^+}+\frac{2i(k+1
)}{g}}}}\prod_{i=1}^{K^{\rm{III}}}{\textstyle{\frac{w_i-x^+-\frac{1}{x^+
}+\frac{i(2k+3)}{g}}{w_i-x^+-\frac{1}{x^+}+\frac{i(2k+1)}{g}}}}\right\}.
\nonumber
\end{eqnarray}
Here the twist $e^{i\alpha}$  enters only the last line.
Eigenvalues are parametrized by solutions of the auxiliary Bethe
equations:
\begin{eqnarray}
\label{bennote}
\prod_{i=1}^{K^{\rm{I}}}\frac{y_k-x^-_i}{y_k-x^+_i}\sqrt{\frac{x^+_i}{x^
-_i}}&=&e^{i\alpha}
\prod_{i=1}^{K^{\rm{III}}}\frac{w_i-y_k-\frac{1}{y_k}-\frac{i}{g}}{w_i-y
_k-\frac{1}{y_k}+\frac{i}{g}},\\
\prod_{i=1}^{K^{\rm{II}}}\frac{w_k-y_i-\frac{1}{y_i}+\frac{i}{g}}{w_k-y_
i-\frac{1}{y_i}-\frac{i}{g}} &=& e^{2i\alpha}\prod_{i=1,i\neq
k}^{K^{\rm{III}}}\frac{w_k-w_i+\frac{2i}{g}}{w_k-w_i-\frac{2i}{g}}.\nonumber
\end{eqnarray}
In the formulae above the variable
$$
v=x^++\frac{1}{x^+}-\frac{i}{g}Q=x^-+\frac{1}{x^-}+\frac{i}{g}Q\,
$$
takes values in the mirror theory rapidity plane, i.e. $x^\pm = x(v
\pm {i\ov g}Q)$ where $x(v)$ is the mirror theory $x$-function. As
was mentioned above, $u_j$ take values in string theory $u$-plane,
and therefore $x_j^\pm = x_s(u_j \pm {i\ov g})$ where $x_s(u)$ is
the string theory $x$-function. These two functions are given by
\begin{align}
&x(u) = \frac{1}{2}(u-i\sqrt{4-u^2}), && x_s(u) = \frac{u}{2}(1+\sqrt{1-\frac{4}{u^2}}).
\end{align}
Finally, the quantities {\small
$\lambda_{\pm}$ are
\begin{eqnarray}\nonumber \hspace{-1cm}
\lambda_\pm(v,u_i,k)&=&\frac{1}{2}\left[1-\frac{(x^-_ix^+-1)
  (x^+-x^+_i)}{(x^-_i-x^+)
  (x^+x^+_i-1)}+\frac{2ik}{g}\frac{x^+
  (x^-_i+x^+_i)}{(x^-_i-x^+)
  (x^+x^+_i-1)}\right.\\ \label{eqn;lambda-pm}
&&~~~~~~~~~~~~\left.\pm\frac{i x^+
  (x^-_i-x^+_i)}{(x^-_i-x^+)
 (x^+x^+_i-1)}\sqrt{4-\left(v-\frac{i(2k-Q)}{g}\right)^2}\right]\,
 .
\end{eqnarray}
}

For a single string theory particle with rapidity $u$ the
 twisted $\mathfrak{sl}(2)$ transfer matrix is of the form {\small
\begin{align}
T^{\alg{sl}(2)}_{Q,1}(v|u) = & 1 + \frac{\left(x^--x^-_1\right) \left(1-x^+_1
x^-\right)}{\left(x^+ - x^-_1\right) \left(1-x^+_1 x^+\right)}
\frac{ x^+}{x^-}
  - 2\cos\alpha \frac{\left(x^+ - x^+_1\right)}{\left(x^+-x^-_1\right)} \sqrt{\frac{x^-_1}{x^+_1}}
   \frac{Q (u - v)}{(u - v)-\frac{i}{g} (Q-1)} + \lambda\, ,
\end{align}
} where {\small
\begin{align*}
\lambda = & (Q - 1) \left(1 +  \frac{(x^+ - x^+_1) (x^-_1 x^+ -
1)}{(x^+ - x^-_1) (x^+_1 x^+ - 1)}\right) + \frac{i}{g} Q (Q -
1)\frac{x^+ (x^-_1 + x^+_1) }{ (x^-_1 - x^+) (x^+_1 x^+ - 1) }\,
\end{align*}
} The matrix $T^{\alg{sl}(2)}_{1,1}(u|v)$ is normalized as
$$
T^{\alg{sl}(2)}_{1,1}(u_*|u) = 1,
$$
which is immediately clear form its expression above. Here the
star indicates analytic continuation to the kinematic region of
string theory.

%
%
 The S-matrix in the string-mirror region $S_{\sl(2)}^{1_*Q}$ is
found in \cite{Arutyunov:2009kf} (see also \cite{BJ09}) and it has
the following weak-coupling expansion
$$
S_{\sl(2)}^{1_*Q}(u,v)=S_0(u,v)+g^2 S_{2}(u,v)+\ldots \, ,
$$
where {\small
\begin{align}
S_0(u,v) = -\frac{\big[(v-u)^2+(Q+1)^2\big]\big[Q-1 + i (v-u)
)\big]}{(u-i)^2 \big[Q-1-i( v-u)\big]}.
\end{align}
} and {\small
\begin{align}
S_2(u,v)& =-S_0(v,u)
\frac{2\big[2Q(u-i)+(u+i)(v^2+Q^2+2v(u-i))\big]}{(v^2+Q^2)
(1+u^2)}+\\
\nonumber &
\frac{S_0(v,u)}{1+u^2}\Big[4\gamma+\psi\left(1+\frac{Q+iv}{2}\right)+\psi\left(1-\frac{Q+iv}{2}\right)+\psi\left(1+\frac{Q-iv}{2}\right)+
\psi\left(1-\frac{Q-iv}{2}\right)\Big]\, .
\end{align}
}These expressions are enough to build up the two leading terms in
the weak-coupling expansion of the asymptotic function $Y^o_Q$.

\subsection{Appendix B: Asymptotic critical points and TBA }

As mentioned in the main text,  a single magnon shows a critical
behaviour quite similar to what has been observed in the
two-particle case in \cite{Arutyunov:2009ax}. However, we would
like to stress that the discussion of criticality is restricted to
the asymptotic solution; the data recently obtained for the
two-particle case \cite{Frolov:2010wt} is not sufficient to
conclusively establish the existence of critical points for an
exact solution.

\smallskip

For model I we find critical behaviour that arises from the
movement of zeroes of $Y_+ - 1$ and poles of $Y_{M|w}$ and
$Y_{M|vw}$ in the complex plane. These zeroes and poles run into
the integration contour at certain values of $g$ and further cross
it, yielding extra contributions in the TBA equations above these
what are called critical values of $g$.

\smallskip

\begin{figure}[h]
\begin{center}
\includegraphics[bb=0 0 300 185,width=3in]{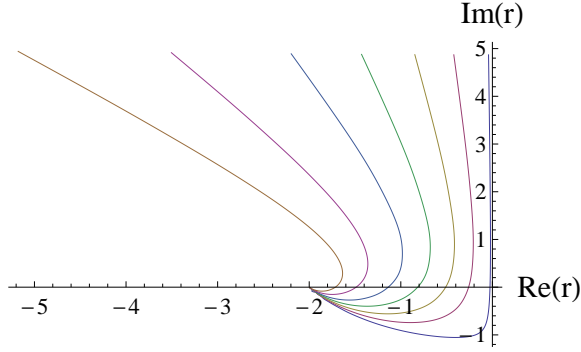}
\end{center}
\caption{The behaviour of zeroes of $Y_+ - 1$ as $g$ increases,
for $p$ taking the values $\frac{9999 \pi}{10000}$, $\frac{3
\pi}{4}$, $\frac{\pi}{2}$, $\frac{\pi}{3}$, $\frac{\pi}{5}$,
$\frac{\pi}{10}$, $\frac{\pi}{100}$ from left to right. For
negative values of momenta the picture should naturally be
mirrored with respect to the imaginary axis.}
\label{fig:Yprootbehav}
\end{figure}

\begin{figure}[h]
\begin{center}
\includegraphics[width=5in]{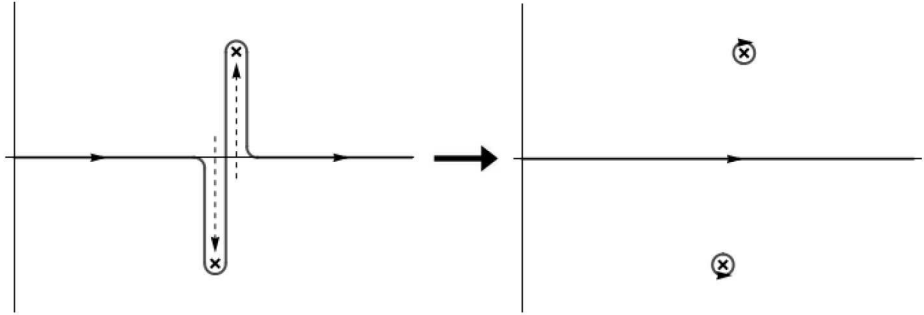}
\end{center}
\caption{The zeroes of $Y_+$ crossing the real line giving extra
contributions as the integration contour is returned to the real
line.} \label{fig:Yproots}
\end{figure}

\begin{figure}[h]
\begin{center}
\hspace{40pt}\includegraphics[width=4in]{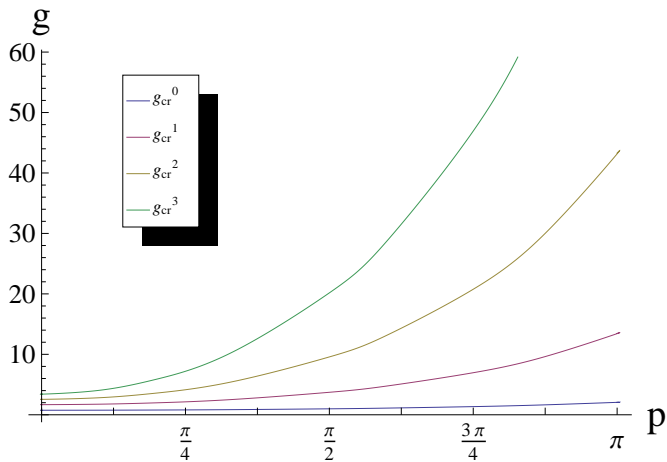}
\end{center}
\caption{The critical values $g_{cr}^M$ as $p$ varies from $0$ to
$\pi$.} \label{fig:gcrit}
\end{figure}

We indicated before that zeroes of $1- Y_+$ and poles of $Y_{M|w}$
and $Y_{M|vw}$ are related by shifts of $i/g$; owing to this we
can treat their critical behaviour in a unified manner. For
concreteness however, let us first consider zeroes of $Y_+ - 1$.
The function $Y_+$ has the interval $(-2,2)$ as its associated
canonical integration contour, and as is plotted in Figure
\ref{fig:Yprootbehav} the  zeroes of $Y_+ - 1$ cross this contour,
and do so at a particular value of $g$ denoted $g_{cr}^{0}$.
As these zeroes do so from either side of the real line they drag
the integration contour along with them. In order to maintain the
canonical integration contour in the TBA equations, one needs to
pick up a contribution from each of these zeroes as they move back
into the complex plane, as illustrated in Figure
\ref{fig:Yproots}.
For the poles of $Y_{M|w}$ and $Y_{M|vw}$ the story is nearly
identical as they show exactly the same behaviour, except that the
crossing happens at higher values in $g$. The critical values as a
function of the particle momentum $p$ have been plotted for the
first few $M$ in Figure \ref{fig:gcrit}.
The contributions these zeroes give in the TBA above the
critical values can straightforwardly be read of from the ground
state TBA equations taking into account the orientation of the
integration contour.

\smallskip

In model II the story is different as $1- Y_+$ has no zeroes, but
now the zeroes $r_M$ show criticality. As discussed in the main
text, we have found exact expressions for the asymptotic zeroes,
allowing us to understand and treat the asymptotic criticality
exactly. Concisely the story is as follows; a zero $r_M$ can be
imaginary but between the real line and $-i/g$ at small coupling
for a given state (momentum). When this is the case, the zero
should be accounted in the excited TBA equations, but as $g$ is
increased the zero moves away from the real line and crosses
either the real line or $-i/g$. The value of $g$ for which this
happens can also be called a critical value, where now beyond the
critical value a zero $r_M$ no longer contributes to the TBA
equations. This story can be readily verified by making use of the
explicit expression (\ref{eq:rvw}). Please note that in this twisted
model, for a carefully chosen state \emph{these critical values
can be arbitrarily small}.

\subsection{Appendix C: Wrapping corrections}

In this section we present the LO corrections to the energy and momentum in full generality for both model I and II.

\subsubsection{Model I}

We write $\psi^{(m)}$ for the $m$-th polygamma function,
$\psi^{(m)} = \frac{d^m \psi^{(0)}(z)}{dz^m}$, where $\psi\equiv
\psi^0(z) = \Gamma'(z)/\Gamma(z)$.
Let us first define
\begin{align}
&E_{\rm LO}^{\rm I} = \frac{(u-\sqrt{u^2+1})^2}{2\pi}\mathscr{E}.
\end{align}
\renewcommand{\baselinestretch}{1.3}
\noindent Then, we introduce the following quantities
{\small
\begin{align}
&A(J)= J\left[\psi^{(J-1)}\left(\frac{1-iu}{2}\right)-(-1)^{J}
\psi ^{(J-1)}\left(\frac{1+i u}{2}\right)\right] \, ,\\
&B_m(J)=\frac{\Gamma (J-m) \Gamma
\left(J-m+\frac{1}{2}\right)}{\Gamma (J-2 (m-1))}\left[\frac{m-1+i
(J-m) u}{(1+iu)^{2m} } + \frac{m-1-i (J-m) u}{(1-iu)^{2m}
}\right].
\end{align}
} \noindent The full leading order wrapping correction to the energy for $J>4$ is given by

{\small
\begin{align}\label{eqn;ExactWrappingE}
\frac{\mathscr{E}(J)}{g^{2J}} =&~ \frac{\pi  \left[(i u+1)^J-(i
u-1)^J\right]}{ 2^{J}(J-1)!\left(u^2+1\right)^{J-1}}
\left\{\frac{A(J)}{4J} +
\frac{i u}{u^2+1}A(J-1) - \frac{J-1}{u^2+1} A(J-2)\right\} + \nonumber\\
&~ -\frac{\sqrt{\pi }}{(J-1)!}\sum_{m=1}^{\left\lfloor
\frac{J}{2}\right\rfloor+1}B_m(J) \zeta (2 J-2 m-1).
\end{align}
}
\renewcommand{\baselinestretch}{1.3}

  \noindent
Formula (\ref{eqn;ExactWrappingE}) was checked analytically up to
$J=8$ and numerically up to $J=200$. It is also readily seen that
the energy correction is real. For $J=3,4$
formula (\ref{eqn;ExactWrappingE}) is not well defined due to the
divergence $\zeta(1)=\infty$. This can be resolved by taking the
appropriate values of $J,n$ and substituting $\zeta(1)\to
\gamma_E$ in (\ref{eqn;ExactWrappingE}). After this substitution,
formula (\ref{eqn;ExactWrappingE}) holds. Explicitly, for $J=3$
one gets {\small
\begin{align}
\frac{\mathscr{E}(3)}{g^6}=& \frac{\pi  \left[(i u+1)^3-(i
u-1)^3\right]}{ 16\left(u^2+1\right)^{2}} \left\{\frac{A(3)}{12} +
\frac{i u}{u^2+1}A(2) - \frac{2}{u^2+1} A(1)\right\} + \nonumber \\
& + \frac{\pi}{2} \frac{(1-3 u^2)\gamma_E + 3 u^2(1 + u^2) \zeta
(3)}{(u^2+1)^3}.
\end{align}
}For $J=2$ one finds ($A(0)=0$) {\small
\begin{align}
\frac{\mathscr{E}(2)}{g^4}=& \frac{\pi  \left[(i u+1)^2-(i
u-1)^2\right]}{4\left(u^2+1\right)} \left\{\frac{A(2)}{8} +
\frac{i u}{u^2+1}A(1) \right\} + \frac{\pi}{2}\frac{(1-4 \gamma_E )u^2-1}{\left(u^2+1\right)^2}.
\end{align}
}

 \noindent Finally, for $J=1$ the sum over $M$ does not converge
and we find that the wrapping correction diverges. Nevertheless,
expression (\ref{eqn;ExactWrappingE}) can be continued to $J=1$
and it gives a finite result {\small
\begin{align}
\frac{\mathscr{E}(1)}{g^2} = -\frac{\pi}{12} \left[ \frac{4
u^2}{\left(u^2+1\right)^2}+3 \psi\left(\frac{1-iu}{2}\right)+3
\psi \left(\frac{1+ i u}{2}\right)\right].
\end{align}
}

So far the formulae for the energy correction were obtained for
arbitrary values of particle rapidity $u$, {\it i.e.} not
necessarily satisfying the quantization condition. If we impose
the quantization condition (the Bethe equations), then at leading
order in $g$ the rescaled rapidity is $u=\cot\frac{\pi n}{J}$ with
$n$ integer. For these values of $u$ the following relation holds \bea
\label{funny} (iu+1)^J-(iu-1)^J=\frac{2^J}{\big(1-e^{\frac{2\pi i
n}{J}}\big)^J}(1-e^{2\pi i n})=0\, . \eea Therefore, the
contribution of the polygamma-functions to the energy correction
completely drops out on solutions of the Bethe equation, resulting in formula (\ref{eqn;ExactWrappingEonshell}).

\subsubsection{Model II}

\subsection*{Energy}
To present the value of the energy integral (\ref{eqn;Luscher})
for model II, we define the following quantities
\renewcommand{\baselinestretch}{1.3} \noindent
{\small
\begin{align}
&A(J)= \left[\psi^{(J)}\left(\frac{1-iu}{2}\right)+(-1)^{J}
\psi ^{(J)}\left(\frac{1+i u}{2}\right)\right] \, ,\\
&B_m(J)= \frac{\Gamma (J-m+1) \Gamma
\left(J-m+\frac{3}{2}\right)}{\Gamma (J-2 (m-1))} \left[\frac{m+i
(J-m+1) u}{(1+iu)^{2m} } + \frac{m-i (J-m+1) u}{(1-iu)^{2m}
}\right].
\end{align}
}
\renewcommand{\baselinestretch}{1.3}

\noindent Then we find for $J>2$ {\small
\begin{align}\label{eqn;ExactWrappingEII}
\frac{E^{\rm II }_{\rm LO}(J)}{g^{2J+4}} =&~ \frac{(i u-1)^J-(i
u+1)^J}{ 2^{J}J!\left(u^2+1\right)^{J+1}}
\left\{\frac{A(J)}{2(J+1)} +
\frac{2 i u}{u^2+1}A(J) - \frac{2J}{u^2+1} A(J-1)\right\} + \\
&~ + \frac{8}{(1+u^2)^2\sqrt{\pi }(J+1)!}\sum_{m=1}^{\left\lfloor
\frac{J}{2}\right\rfloor+1}B_m(J) \zeta (2 J-2 m + 1) -
\frac{8}{(1+u^2)^2}\frac{\Gamma(J+\frac{3}{2})}{\Gamma(J+2)}\frac{\zeta(2J+1)}{\sqrt{\pi}}.\nonumber
\end{align}
}Formula (\ref{eqn;ExactWrappingEII}) was checked analytically up
to $J=8$ and numerically up to $J=300$. For $J=2$ formula
(\ref{eqn;ExactWrappingE}) is not well defined, again because of the divergence
$\zeta(1)=\infty$. However, for $J=2$, the energy correction can be easily evaluated
directly {\small
\begin{align}
\frac{E^{\rm II }_{\rm LO}(2)}{g^8}=& \frac{(i u-1)^2-(i u+1)^2}{
8\left(u^2+1\right)^{3}} \left\{\frac{A(3)}{6} + \frac{2 i
uA(2)}{u^2+1} - \frac{4A(1)}{u^2+1} \right\} + \nonumber
\frac{\left(6 u^2+2\right) \zeta
(3)}{\left(u^2+1\right)^4}-\frac{5 \zeta (5)}{2
\left(u^2+1\right)^2}.
\end{align}
}Finally, for $J=1$ the sum over $M$ does not converge and,
therefore, the leading wrapping correction diverges.

\smallskip

\subsection*{Momentum}

Define {\small
\begin{align}
\nonumber D_m(J)=&\, i\frac{
2^{2m-2J+2}\Gamma(2J-2m+2)}{\Gamma(J+3)\Gamma(J-2m+2)}\left[
\frac{(J-m+1)(2m+1)}{1+u^2}\left(\frac{1}{(1+iu)^{2m+2}}-\frac{1}{(1-iu)^{2m+2}}\right)+\right.\\
&\left.+\frac{2(J+2)m}{(1+u^2)^2}\left(\frac{1}{(1+iu)^{2m+1}}-\frac{1}{(1-iu)^{2m+1}}\right)
\right],
\end{align}
}then {\small
\begin{align}
&P^{\mathrm{II}}_{\rm LO} =-\sum_{m=0}^{\lfloor\frac{J+1}{2}\rfloor}D_m(J)\zeta(2J-2m+1)+\nonumber\\
&+ i\frac{(i u-1)^J-(i u+1)^J}{
2^{J+2}(J+2)!\left(u^2+1\right)^{J+1}}A(J+2) +
 i\frac{(i u+1)(i u-1)^{J}-(i u-1)(i u+1)^{J}}{ 2^{J+1}(J+1)!\left(u^2+1\right)^{J+2}}A(J+1) \nonumber\\
&-i \frac{(i u-1)^{J+2}-(i u+1)^{J+2}}{
2^{J}J!\left(u^2+1\right)^{J+3}}A(J) -i\frac{(i u+1)(i
u-1)^{J+2}-(i u-1)(i u+1)^{J+2}}{
2^{J-1}(J-1)!\left(u^2+1\right)^{J+4}}A(J-1)\, . \nonumber
\end{align}}
On the solution of the Bethe equations the polygamma functions do not drop out.

\subsection{Appendix D: Details on the NLO computation}

In this appendix we have gathered some details on the explicit computation of the NLO correction to the energy.

\subsubsection{The Harmonic number integral}\label{app;HarmonicIntegral}
Special attention has to be paid to the contribution $Y^{\rm NLO
}_{\psi}$ due to the appearance of harmonic numbers. Let us first
set up the tools to handle sums and integrals of these numbers. The
definition of the multiple (double in this case) zeta value
$\zeta(a,b)$ is given by
\begin{align}
\zeta(a,b)\equiv \sum_{n>m}\frac{1}{n^a}\frac{1}{m^b}.
\end{align}
For integer values of $a,b$ this can generically be written in
terms of regular $\zeta$ values and products thereof. A useful
program on the web for doing this is EZFace{\footnote{http://oldweb.cecm.sfu.ca/projects/EZFace/}}. These multiple zeta
values can then be related to sums of polygamma function via
harmonic sums \cite{Kotikov:2005gr}
\begin{align}\label{eqn;sumPolyGammaA}
\sum_{Q=1}^{\infty} \frac{\psi^{(b)}(Q)}{Q^a} = (-1)^{b+1} b!
[\zeta (a) \zeta (b+1)-\zeta(a,b+1)].
\end{align}
We will also encounter more generic sums including
$\psi$-functions. Following \cite{BJ09} we can write for sums of
polygamma's
\begin{align}\label{eqn;sumPolyGammaB}
\sum_{Q=1}^{\infty}\Sigma(Q)\psi^{(b)}(Q) =(-1)^b\int_0^\infty dt
\mathcal{L}^{-1}(\Sigma)\Gamma(b+1)\frac{\mathrm{Li}_{b+1}(e^{-t})-\zeta(b+1)}{e^{t}-1},
\end{align}
where $\mathcal{L}^{-1}(\Sigma)$ is the inverse Laplace transform
of $\Sigma(Q)$. For the digamma function one finds
\begin{align}\label{eqn;sumPolyGammaC}
\sum_{Q=1}^{\infty}\Sigma(Q)\psi^{(0)}(Q) =\int_0^\infty dt
\mathcal{L}^{-1}(\Sigma)\frac{\log \left(1-e^{-t}\right)+\gamma
}{1-e^t}.
\end{align}
In general when one encounters rational functions
$\sum_{Q=1}^{\infty}\Sigma(Q)\psi^{(b)}(Q)$, the recipe that
proved useful in explicit computations was to split it into two
parts $\Sigma(Q) = \Sigma_0(Q)+\Sigma_1(Q)$, where
\begin{align}
\Sigma_0(Q) = \sum_{n>1} \frac{a_n}{Q^n}.
\end{align}
To the term containing $\Sigma_0$ one can then apply formula
(\ref{eqn;sumPolyGammaA}) and on the remainder $\Sigma_1$ one can
use the integral relations (\ref{eqn;sumPolyGammaB}) and
(\ref{eqn;sumPolyGammaC}).

\smallskip

Let us now apply the above discussion to the integral $E_{\rm NLO}^{(4)}$.
The integral is most easily evaluated by summing residues. The
integrand has three different types of poles in the upper
half-plane
\begin{itemize}
 \item dynamical poles at $u+i(Q\pm1)$, giving a contribution $E_{{\rm NLO},dyn}^{(4)}$
 \item poles at $i(Q+2k)$ for $k\in\mathbb{Z}_{>0}$, giving $E_{{\rm NLO},k}^{(4)}$
 \item pole at $iQ$, resulting in $E_{{\rm NLO},iQ}^{(4)}$
\end{itemize}
All of the above poles give rise to residues that contribute to
the integral and we will discuss them separately.

\subsection*{Dynamical poles}

Let us first consider the residue at the points $v=u+i(Q\pm1)$.
One straightforwardly finds
\begin{align}
\mathrm{res}_{v=u+i(Q+1)}Y^{\rm NLO}_\psi &= \frac{8 i Q (Q+1)
\left(H_{Q-\frac{i u-1}{2}}+H_{\frac{i
u-1}{2}-Q}+H_{\frac{1-iu}{2}}+H_{\frac{
 (iu-1)}{2}}\right)}{(u+i)^J (u^2+1)^3 (2 i Q+u+i)^2 (2 iQ+u+i)^{J}},\\
\mathrm{res}_{v=u+i(Q-1)}Y^{\rm NLO}_\psi &= \frac{8 i Q(Q-1)
\left(H_{Q-\frac{i u+1}{2}}+H_{\frac{i u+1}{2}-Q}+H_{-\frac{i
u+1}{2}} + H_{\frac{i u+1}{2}}\right)}{(u-i)^J(u^2+1)^3(-2 Q+i
u+1)^2(2 i Q+u-i)^{J}}.
\end{align}
Both contributions have to be added and then summed over $Q$. The
obtained expression naturally splits in two pieces, one which is
purely rational and one that also contains terms of the form
$\psi(Q+\frac{1+iw}{2})$. The latter can be summed by making use
of the techniques of \cite{BJ09} used to derive equations
(\ref{eqn;sumPolyGammaA}-\ref{eqn;sumPolyGammaC}). This yields the
following result for the first piece
\begin{align}
E_{{\rm NLO},dyn}^{(4),1}=\frac{i (-1)^J
(u+i)^J}{(u^2+1)^{J+3}2^{J-1}} \left[\frac{\psi
^{(J)}\left(\frac{1-iu}{2}\right)}{J!} -
\frac{\left(u^2+1\right)}{4}
\frac{\psi^{(J+2)}\left(\frac{1-iu}{2}\right)}{(J+2)!} - iu
\frac{\psi ^{(J+1)}\left(\frac{1-iu}{2}\right)}{(J+1)!}\right].
\end{align}
The remaining rational part can be more straightforwardly summed.
Adding up both contributions gives
\begin{align}
E_{{\rm NLO},dyn}^{(4)}=\frac{-8i^{J} u}{2^{J}(u^2+1)^4 (u-i)^J }
\left[\frac{u^2+1}{4} \frac{\psi
^{(J+1)}\left(\frac{1-iu}{2}\right)}{(J+1)!} + i u \frac{\psi
^{(J)}\left(\frac{1-iu}{2}\right)}{J!} - \frac{ \psi
^{(J-1)}\left(\frac{1-iu}{2}\right)}{(J-1)!}\right].
\end{align}
In deriving the above results we used the fact that the Bethe root
$u$ satisfies the Bethe equations
$\left(\frac{u-i}{u+i}\right)^J=1$. As it is an intermediate
result, the quantity $E_{{\rm NLO},dyn}^{(4)}$ is not real by
itself. It will be canceled in the final energy correction as it should because it solely consists of $\psi$-functions.

\subsection*{Poles at $i(Q+2k)$}

The integrand also exhibits a pole at $i(Q+2k)$ due to the
harmonic numbers. The residue at this point is easily found to be
\begin{align}
\mathrm{res}_{i(Q+2k)} Y^{\rm NLO}_{\psi} = 4i\, Y^{\rm
LO}_Q(i(Q+2k),u).
\end{align}
In other words, this gives a total contribution of
\begin{align}
E_{{\rm NLO},k}^{(4)} = 4\sum_{k=1}^{\infty}\sum_{Q=1}^{\infty}
Y^{\rm LO}_Q(i(Q+2k),u).
\end{align}
It is most convenient to first preform the sum over $Q$. This
yields, again on a solution of the Bethe equation,
\begin{align}
&\sum_{Q=1}^{\infty} Y^{\rm LO}_Q(i(Q+2k),u) =\\
&\frac{2^{1-J}(-i)^J}{(u+i)^{J} \left(u^2+1\right)^3} \frac{u+i-2
i k}{k^{J+2}}
+ \sum_{n=0}^{J-2}B^{\psi}_{n}(J)-\frac{2^{3-2 J}i}{\left(u^2+1\right)^3}\frac{\psi^{(J)}(k+1)}{k^{J+1}J!}+ \nonumber\\
&-\frac{2^{4-2 J}}{(u^2+1)^3}\frac{u}{k^{J} [(2 k+i u)^2-1]}\left[
\frac{(u^2+1)\psi^{(J+1)}(k+1)}{4iu(J+1)!} + \frac{\psi
^{(J)}(k+1)}{J!}-\frac{\psi
^{(J-1)}(k+1)}{iu(J-1)!}\right]\nonumber
\end{align}
where
\begin{align}
B^{\psi}_{n}(J)= \frac{(-i)^n}{ 2^{2 J-n-1}}
\left[\frac{1}{(u+i)^{n+1}}-\frac{1}{(u-i)^{n+1}}\right]
\left[\frac{u^2+1}{k^{J+2}}-\frac{4 i u}{k^{J+1}}-\frac{4}{
k^{J}}\right] \frac{ \psi^{(J-n-1)}(k+1)}{\left(u^2+1\right)^3
\Gamma (J-n)}.
\end{align}
By using the relation
\begin{align}
\psi ^{(n)}(k+1)=\psi ^{(n)}(k)+\frac{(-1)^n n!}{ k^{n+1}},
\end{align}
it is now straightforward to sum first line over $k$ by means of
the previously discussed techniques.

\smallskip

The terms from the second line are all proportional to
$\frac{u}{k^{J} [(2 k+i u)^2-1]}$. This function can easily be
split up in the following way
\begin{align}
\frac{1}{k^{J} [(2 k+i u)^2-1]} =\frac{(2 i)^J}{(u+i)^{J}}
\frac{1}{(2 k+i u)^2-1} + \sum_{n=0}^{J-2}
\left[\frac{1}{(u+i)^{n}}-\frac{1}{(u-i)^{n}}\right]\frac{(2i)^{n-2}}{k^{J-n}}.
\end{align}
The tail allows for a direct expression in terms of multiple zeta
values. The first piece has to be inverse Laplace transformed
\begin{align}
\mathcal{L}^{-1}\left(\frac{1}{(2 k+i u)^2-1}\right) =
\frac{1}{4}\left(e^t-1\right) e^{-\frac{1}{2} t (1+i u)}.
\end{align}
One can then explicitly preform the integral
(\ref{eqn;sumPolyGammaB}) to find
\begin{align}
&\int dt \mathcal{L}^{-1}\left(\frac{1}{(2 k+i u)^2-1}\right)\Gamma(k+1)\frac{\mathrm{Li}_{k+1}(e^{-t})-\zeta(k+1)}{e^{t}-1} =\\
&\qquad\quad \frac{2 i (-1)^k k! \zeta (k+1)}{u-i}-(-1)^k (2
i)^{k+1} k! \frac{H_{\frac{1}{2} i (u+i)}}{(u+i)^{k+1}} - (-1)^k
k! \sum_{n=1}^{k} \frac{(2 i)^n \zeta
(k-n+2)}{(u+i)^{n}}.\nonumber
\end{align}

\subsection*{Pole at $iQ$}

The residue at $v=iQ$ is the most complicated one. It receives
contributions from both the rational part and the harmonic number
part. Let us expand the harmonic number part around this
point
\begin{align}
&H_{-\frac{Q+iv}{2}} + H_{\frac{i v-Q}{2}}+H_{\frac{Q-i v}{2}} + H_{\frac{Q+i v}{2}} = \\
&\frac{2i}{v-iQ} + (H_Q+H_{Q-1}) -\sum_{n=2}
\left(\frac{1}{2i}\right)^n \frac{2 (-1)^{n+1} \psi ^{(n)}(1)
-\psi ^{(n)}(Q)-\psi ^{(n)}(Q+1)}{n!}(v-iQ)^n.\nonumber
\end{align}
For any value of $J$ one can now easily extract the residue at
this point and apply the discussed techniques.

\subsubsection{Some integrals}\label{section;Integrals}

In this section we present a list with integrals that are
encountered throughout the main text. These integrals are
\begin{align}
&\mathcal{I}_1 = \frac{1}{2\pi}\sum_{Q=1}^{\infty} \int dv ~ \frac{v-u}{Q^2-1+v^2-u^2}Y^{o}_{Q}(v)\, ,\\
&\mathcal{I}_2 = \frac{(u^2+1)^2}{2\pi}\sum_{Q=1}^{\infty}\int dv~\frac{4 v^2
}{\left(Q^2+v^2\right)^2}Y^{o}_Q(v) \, ,\\
&\mathcal{I}_3 =\frac{1}{2\pi}\sum_{Q=1}^{\infty}\int dv~
\frac{(Q^2-1)}{(Q^2+v^2)(1-Q^2-v^2+u^2)}Y^{o}_Q(v)\, .
\end{align}
Integral $\mathcal{I}_1$ is computed similarly to $E_{\rm LO}$.
One finds
\begin{align}
2\pi \mathcal{I}_1(J)=\frac{A_{\mathcal{I}}(J+1)+B_{\mathcal{I}}(J+1)}{(u^2+1)^2(J+1)!}
-\frac{A_{\mathcal{I}}(J+2)+B_{\mathcal{I}}(J+2)}{(u^2+1)(J+2)!},
\end{align}
where the coefficients in the above expression are
\begin{align}
\frac{A_{\mathcal{I}}(J)}{2\pi i}=&
\frac{J}{2^J} \left[\frac{1}{(u+i)^{J-1}}-\frac{1}{(u-i)^{J-1}}\right]\left[\psi ^{(J-1)}\left(\frac{1-iu}{2}\right)-(-1)^J \psi ^{(J-1)}\left(\frac{1+iu}{2}\right)\right] + \\
& + \frac{J(J-1)}{2^{J-1}}
\left[\frac{1}{(u+i)^{J}}-\frac{1}{(u-i)^{J}}\right]
   \left[\psi ^{(J-2)}\left(\frac{1-iu}{2}\right)+(-1)^{J} \psi ^{(J-2)}\left(\frac{1+iu}{2}\right)\right],\nonumber\\
\frac{B_{\mathcal{I}}(J)}{2\pi i}&=
\frac{2J}{\sqrt{\pi}}\sum_{n=1}^{\lfloor\frac{J}{2}\rfloor}
\frac{\Gamma (J-n) \Gamma\left(J-n+\frac{1}{2}\right)}{\Gamma (J-2
n+1)} \left[\frac{1}{(u-i)^{2n}}-\frac{1}{(u+i)^{2n}}\right] \zeta
(2 J-2 n-1).
\end{align}
One can then express the remaining integrals $\mathcal{I}_{2,3}$
in terms of the previously computed integrals and derivatives thereof
\begin{align}
&\mathcal{I}_2(J) = -\frac{32 E^{\mathrm{I}}_{\rm LO
}(J+2)}{(J+2)(J+3)}
-\frac{d}{du}\frac{(u^2+1)^2 P^{\mathrm{II}}_{\rm LO}(J)}{J+3}+\nonumber\\
&\qquad\qquad+\frac{d}{du}\frac{4\left(u^2+1\right)^2\mathcal{I}_1(J)}{(J+2)(J+3)}
+
\frac{2\left(u^2+1\right)^2 E^{\mathrm{II}}_{\rm LO}(J+1)}{J+3}\, ,\\
&\mathcal{I}_3(J)=u \mathcal{I}_1(J+1) -E_{\rm LO
}^{\mathrm{II}}(J+1)-\frac{4 E_{\rm LO}^{\mathrm{I}}(J+2) +
2\partial_u[ \mathcal{I}_1(J)(u^2+1)]}{ (J+2)(u^2+1)^2}\, .
\end{align}


\end{document}